\documentclass[preprint,superscriptaddress,unsortedaddress,showpacs,
preprintnumbers,amsmath,amssymb,nofootinbib]{revtex4}

\usepackage{hyperref}
\usepackage{graphicx}
\usepackage{dcolumn}
\usepackage{bm}
\usepackage{color}

\begin{document}


\title{The Role of Mesons in the Electromagnetic\\
Form Factors of the Nucleon}


\author{C.~Crawford}
\affiliation{University of Kentucky, Lexington, KY 40504}

\author{T. Akdogan}
\affiliation{Physics Department and/or Laboratory for Nuclear Science
and/or Center for Theoretical Physics and/or Bates Linear Accelerator
Center, Massachusetts Institute of Technology, Cambridge, MA 02139}

\author{R. Alarcon}
\affiliation{Arizona State University, Tempe, AZ 85287}

\author{W. Bertozzi}
\affiliation{Physics Department and/or Laboratory for Nuclear Science
and/or Center for Theoretical Physics and/or Bates Linear Accelerator
Center, Massachusetts Institute of Technology, Cambridge, MA 02139}

\author{E. Booth}
\affiliation{Boston University, Boston, MA 02215}

\author{T. Botto}
\affiliation{Physics Department and/or Laboratory for Nuclear Science
and/or Center for Theoretical Physics and/or Bates Linear Accelerator
Center, Massachusetts Institute of Technology, Cambridge, MA 02139}

\author{J.R. Calarco}
\affiliation{University of New Hampshire, Durham, NH 03824}

\author{B. Clasie}
\affiliation{Physics Department and/or Laboratory for Nuclear Science
and/or Center for Theoretical Physics and/or Bates Linear Accelerator
Center, Massachusetts Institute of Technology, Cambridge, MA 02139}

\author{A. DeGrush}
\affiliation{Physics Department and/or Laboratory for Nuclear Science
and/or Center for Theoretical Physics and/or Bates Linear Accelerator
Center, Massachusetts Institute of Technology, Cambridge, MA 02139}

\author{T.~W.~Donnelly}
\affiliation{Physics Department and/or Laboratory for Nuclear Science
and/or Center for Theoretical Physics and/or Bates Linear Accelerator
Center, Massachusetts Institute of Technology, Cambridge, MA 02139}

\author{K. Dow}
\affiliation{Physics Department and/or Laboratory for Nuclear Science
and/or Center for Theoretical Physics and/or Bates Linear Accelerator
Center, Massachusetts Institute of Technology, Cambridge, MA 02139}

\author{M. Farkhondeh}
\affiliation{Physics Department and/or Laboratory for Nuclear Science
and/or Center for Theoretical Physics and/or Bates Linear Accelerator
Center, Massachusetts Institute of Technology, Cambridge, MA 02139}

\author{R. Fatemi}
\affiliation{Physics Department and/or Laboratory for Nuclear Science
and/or Center for Theoretical Physics and/or Bates Linear Accelerator
Center, Massachusetts Institute of Technology, Cambridge, MA 02139}

\author{O. Filoti}
\affiliation{University of New Hampshire, Durham, NH 03824}

\author{W. Franklin}
\affiliation{Physics Department and/or Laboratory for Nuclear Science
and/or Center for Theoretical Physics and/or Bates Linear Accelerator
Center, Massachusetts Institute of Technology, Cambridge, MA 02139}

\author{H. Gao}
\affiliation{Duke University, Durham, NC 27708-0305}

\author{E. Geis}
\affiliation{Arizona State University, Tempe, AZ 85287}

\author{S. Gilad}
\affiliation{Physics Department and/or Laboratory for Nuclear Science
and/or Center for Theoretical Physics and/or Bates Linear Accelerator
Center, Massachusetts Institute of Technology, Cambridge, MA 02139}

\author{D. Hasell}
\affiliation{Physics Department and/or Laboratory for Nuclear Science
and/or Center for Theoretical Physics and/or Bates Linear Accelerator
Center, Massachusetts Institute of Technology, Cambridge, MA 02139}

\author{P. Karpius}
\affiliation{University of New Hampshire, Durham, NH 03824}

\author{M. Kohl}
\affiliation{Hampton University, Hampton, VA 23664}

\author{H. Kolster}
\affiliation{Physics Department and/or Laboratory for Nuclear Science
and/or Center for Theoretical Physics and/or Bates Linear Accelerator
Center, Massachusetts Institute of Technology, Cambridge, MA 02139}

\author{T. Lee}
\affiliation{University of New Hampshire, Durham, NH 03824}

\author{E.~Lomon}
\affiliation{Physics Department and/or Laboratory for Nuclear Science
and/or Center for Theoretical Physics and/or Bates Linear Accelerator
Center, Massachusetts Institute of Technology, Cambridge, MA 02139}

\author{A. Maschinot}
\affiliation{Physics Department and/or Laboratory for Nuclear Science
and/or Center for Theoretical Physics and/or Bates Linear Accelerator
Center, Massachusetts Institute of Technology, Cambridge, MA 02139}

\author{J. Matthews}
\affiliation{Physics Department and/or Laboratory for Nuclear Science
and/or Center for Theoretical Physics and/or Bates Linear Accelerator
Center, Massachusetts Institute of Technology, Cambridge, MA 02139}

\author{K. McIlhany}
\affiliation{United States Naval Academy, Annapolis, MD 21402}

\author{N. Meitanis}
\affiliation{Physics Department and/or Laboratory for Nuclear Science
and/or Center for Theoretical Physics and/or Bates Linear Accelerator
Center, Massachusetts Institute of Technology, Cambridge, MA 02139}

\author{R. Milner}
\affiliation{Physics Department and/or Laboratory for Nuclear Science
and/or Center for Theoretical Physics and/or Bates Linear Accelerator
Center, Massachusetts Institute of Technology, Cambridge, MA 02139}

\author{J. Rapaport}
\affiliation{Ohio University, Athens, OH 45701}

\author{R. Redwine}
\affiliation{Physics Department and/or Laboratory for Nuclear Science
and/or Center for Theoretical Physics and/or Bates Linear Accelerator
Center, Massachusetts Institute of Technology, Cambridge, MA 02139}

\author{J. Seely}
\affiliation{Physics Department and/or Laboratory for Nuclear Science
and/or Center for Theoretical Physics and/or Bates Linear Accelerator
Center, Massachusetts Institute of Technology, Cambridge, MA 02139}

\author{A. Shinozaki}
\affiliation{Physics Department and/or Laboratory for Nuclear Science
and/or Center for Theoretical Physics and/or Bates Linear Accelerator
Center, Massachusetts Institute of Technology, Cambridge, MA 02139}

\author{A. Sindile}
\affiliation{University of New Hampshire, Durham, NH 03824}

\author{S. \v{S}irca}
\affiliation{Physics Department and/or Laboratory for Nuclear Science
and/or Center for Theoretical Physics and/or Bates Linear Accelerator
Center, Massachusetts Institute of Technology, Cambridge, MA 02139}

\author{E. Six}
\affiliation{Physics Department and/or Laboratory for Nuclear Science
and/or Center for Theoretical Physics and/or Bates Linear Accelerator
Center, Massachusetts Institute of Technology, Cambridge, MA 02139}

\author{T. Smith}
\affiliation{Dartmouth College, Hanover, NH 03755}

\author{B. Tonguc}
\affiliation{Arizona State University, Tempe, AZ 85287}

\author{C. Tschalaer}
\affiliation{Physics Department and/or Laboratory for Nuclear Science
and/or Center for Theoretical Physics and/or Bates Linear Accelerator
Center, Massachusetts Institute of Technology, Cambridge, MA 02139}

\author{E. Tsentalovich}
\affiliation{Physics Department and/or Laboratory for Nuclear Science
and/or Center for Theoretical Physics and/or Bates Linear Accelerator
Center, Massachusetts Institute of Technology, Cambridge, MA 02139}

\author{W. Turchinetz}
\affiliation{Physics Department and/or Laboratory for Nuclear Science
and/or Center for Theoretical Physics and/or Bates Linear Accelerator
Center, Massachusetts Institute of Technology, Cambridge, MA 02139}

\author{Y. Xiao}
\affiliation{Physics Department and/or Laboratory for Nuclear Science
and/or Center for Theoretical Physics and/or Bates Linear Accelerator
Center, Massachusetts Institute of Technology, Cambridge, MA 02139}

\author{W. Xu}
\affiliation{Duke University, Durham, NC 27708-0305}

\author{C. Zhang}
\affiliation{Physics Department and/or Laboratory for Nuclear Science
and/or Center for Theoretical Physics and/or Bates Linear Accelerator
Center, Massachusetts Institute of Technology, Cambridge, MA 02139}

\author{Z. Zhou}
\affiliation{Physics Department and/or Laboratory for Nuclear Science
and/or Center for Theoretical Physics and/or Bates Linear Accelerator
Center, Massachusetts Institute of Technology, Cambridge, MA 02139}

\author{V. Ziskin}
\affiliation{Physics Department and/or Laboratory for Nuclear Science
and/or Center for Theoretical Physics and/or Bates Linear Accelerator
Center, Massachusetts Institute of Technology, Cambridge, MA 02139}

\author{T. Zwart}
\affiliation{Physics Department and/or Laboratory for Nuclear Science
and/or Center for Theoretical Physics and/or Bates Linear Accelerator
Center, Massachusetts Institute of Technology, Cambridge, MA 02139}


\date{\today}


\begin{abstract}
The roles played by mesons in the electromagnetic form factors of the
nucleon are explored using as a basis a model containing vector mesons
with coupling to the continuum together with the asymptotic $Q^2$
behavior of perturbative QCD. Specifically, the vector dominance model
(GKex) developed by Lomon is employed, as it is known to be very
successful in representing the existing high-quality data published to
date. An analysis is made of the experimental uncertainties present
when the differences between the GKex model and the data are expanded
in orthonormal basis functions. A main motivation for the present
study is to provide insight into how the various ingredients in this
model yield the measured behavior, including discussions of when
dipole form factors are to be expected or not, of which mesons are the
major contributors, for instance, at low-$Q^2$ or large distances, and
of what effects are predicted from coupling to the continuum. Such
insights are first discussed in momentum space, followed by an
analysis of how different and potentially useful information emerges
when both the experimental and theoretical electric form factors are
Fourier transformed to coordinate space.  While these Fourier
transforms should not be interpreted as ``charge distributions",
nevertheless the roles played by the various mesons, especially which
are dominant at large or small distance scales, can be explored via
such experiment--theory comparisons.
\end{abstract}




\maketitle


\section{Introduction}
\label{sec:intro}

Whether one uses hadronic language involving some set of baryons and
mesons or QCD language with quarks and gluons, the nucleon is not a
point Dirac particle, but has spatial extension. Its properties may be
described, in a large part, in terms of a set of elastic electric and
magnetic form factors, $G_E^p$, $G_E^n$, $G_M^p$, and $G_M^n$ arising
in EM elastic electron scattering from protons and neutrons, $G_A$
arising when the weak interaction plays a role, together with $G_E^s$
and $G_M^s$, the strangeness form factors which may play a role in
parity-violating elastic electron scattering. In this paper we focus
on the first four, the electric and magnetic form factors of the
proton and neutron.
Clearly having a detailed understanding of all of the form factors of
the nucleon constitutes a major goal in physics. These are central to
our understanding of strongly-coupled QCD and form the building blocks
for much of what is done in exploring the electroweak structure of
nuclei.

In experimental studies the electromagnetic form factors of the proton
have traditionally been extracted using the Rosenbluth equation for
elastic electron scattering from hydrogen, {\it i.e.} with no
polarization information (no polarized electrons, no polarized
hydrogen target, no measurement of the recoiling proton
polarization). The Rosenbluth differential cross section may be
written as:
\begin{equation}
 \frac{d\sigma_0}{d\Omega}(E_e, \theta_e)=\frac{\sigma_{Mott}(E_e, \theta_e )}
 {(1+\tau)\varepsilon}
 \left[\varepsilon (G_E^p(\tau))^2+\tau(G_M^p(\tau))^2 \right]
 \label{eq:sig0}
\end{equation}
where $\tau\equiv|Q^2|/4m_p^2$ is the dimensionless 4-momentum
transfer and
\begin{equation}
 \varepsilon\equiv\left[1+2(1+\tau)\tan^2{\theta_e/2}\right]^{-1}
 \label{eq:longpol}
\end{equation}
is the so-called virtual photon polarization, governing the balance
between longitudinal and transverse responses. The factor
$\sigma_{Mott}$ is the Mott cross section~\citep{Bjorken:1964aa}, {\it
i.e.} the cross section for scattering from structureless fermions. In
deriving Eq.~(\ref{eq:sig0}) it has been assumed that the
one-photon-exchange approximation is valid. In principle, by varying
the electron scattering angle $\theta_e$ at fixed $\tau$, one can
separate $G_E^p$ from $G_M^p$. At low $Q^2$ this is the usual
procedure; however, at high $Q^2$ typically the term involving the
magnetic form factor dominates, with the term involving the electric
form factor contributing only at the few percent level.

Effects beyond the one-photon-exchange approximation are thought to
play a significant role~\citep{Guichon:2003qm, Blunden:2005ew,
Carlson:2007sp} and thereby modify Eq.~(\ref{eq:sig0}) from its
standard Rosenbluth form. At low $Q^2$ the present understanding is
that such contributions provide relatively small corrections, and thus
Eq.~(\ref{eq:sig0}) is a reasonably good approximation.  In contrast,
at high $Q^2$ this is not believed to be the case, making relatively
large corrections necessary before $G_E^p$ can be extracted using the
Rosenbluth cross section. A simple estimate can help to make this
clear.  Defining the ratio
\begin{equation}
  \xi_p \equiv\frac{G_E^p}{\sqrt{\tau} G_M^p}=
  \frac{R_p}{\mu_p \sqrt{\tau}}
  \label{eq:xi}
\end{equation}
where $R_p \equiv \mu_p G_E^p / G_M^p$ (see discussions in
Sect.~\ref{sec:momspace}), the Rosenbluth cross section in
Eq.~(\ref{eq:sig0}) is seen to be proportional to $1+\varepsilon
\xi_p^2$. Using either the model to be discussed in the next section
or the data in the following section, one finds that at $Q^2=1 (5)$
(GeV/c)$^2$ one has $\xi_p \sim 0.6 (0.1)$.  Accordingly, in the
latter case the second term (the one containing $(G_E^p)^2$) is only
about 1\% of the first term, namely, about $\alpha$; as a
consequence it is not surprising that higher-order QED corrections
play a role. This issue will be definitively resolved when new
measurements are made using both electrons and positrons to exploit
the sign change that occurs in the interference between one and
two-photon exchange contributions when the lepton sign is reversed.
Experiments are planned or in progress to address these
issues at JLab, Novosibirsk and DESY(OLYMPUS).

In recent decades new approaches have been used to separate $G_E^p$
from $G_M^p$, namely by using polarized electrons and either polarized
hydrogen targets, $^1\vec{H}(\vec{e}, e^\prime p)$, or by measuring
the recoil polarization of the proton in the final state after the
elastic scattering, $^1H(\vec{e}, e^\prime\vec{p})$. For instance, for
the polarized electron/polarized target case one has
\begin{equation}
 \frac{{d\sigma}}{{d\Omega}}(E_e, \theta_e;\theta^*, \phi^*)=
 \frac{{d\sigma _0}}{{d\Omega}}(E_e, \theta_e)\left[1+p_e
 \vec{p}_T\cdot\vec{A}(\tau, \varepsilon; \theta^*, \phi^*)\right]
 \label{eq:sigpol}
\end{equation}
where $p_e$ is the longitudinal electron polarization, $\vec{p}_T$ is
a vector pointing in the direction characterized by the angles
$(\theta^*, \phi^*)$ in a coordinate system with z-axis along the
virtual photon direction and with the normal to the electron
scattering plane lying along the y-axis (see~\citep{Donnelly:1985ry}).
The polarization information is contained in the product
\begin{equation}
 \vec{p}_T\cdot\vec{A}(\tau, \varepsilon; \theta^*, \phi^*)\sim
 \sqrt{2\varepsilon} {G_E^p}(\tau){G_M^p}(\tau)\sin{\theta^*}\cos{\phi^*}+
 \sqrt{\tau(1+\varepsilon)}(G_M^p(\tau))^2\cos{\theta^*}
 \label{eq:polpiece}
\end{equation}
and clearly by flipping the electron's helicity and/or the target's
spin and choosing the target polarization to lie in at least two
different directions it is possible, at least in principle, to
separate the interference $G_E^pG_M^p$ from the term having
$(G_M^p)^2$. Experimentally it is clearly advantageous to form a
ratio of the result given above for two choices of polarization
directions, say $(\theta_1^*,\phi_1^*)$ and $(\theta_2^*,\phi_2^*)$:
\begin{eqnarray}
  \frac{\vec{p}_T\cdot\vec{A}(\tau, \varepsilon;
  \theta_1^*, \phi_1^*)}{\vec{p}_T\cdot\vec{A}(\tau, \varepsilon;
  \theta_2^*, \phi_2^*)} &=& \frac{\sqrt{2\varepsilon}
  {G_E^p}(\tau){G_M^p}(\tau)\sin{\theta_1^*}\cos{\phi_1^*}+
  \sqrt{\tau(1+\varepsilon)}(G_M^p(\tau))^2\cos{\theta_1^*}}
  {\sqrt{2\varepsilon}
  {G_E^p}(\tau){G_M^p}(\tau)\sin{\theta_2^*}\cos{\phi_2^*}+
  \sqrt{\tau(1+\varepsilon)}(G_M^p(\tau))^2\cos{\theta_2^*}} \label{eq:pol-a}\\
  &=& \frac
  {\frac{\sqrt{2\varepsilon}}{\sqrt{\tau(1+\varepsilon)}}
  \frac{G_E^p(\tau)}{G_M^p(\tau)}\sin{\theta_1^*}\cos{\phi_1^*}
  +\cos{\theta_1^*} }
  {\frac{\sqrt{2\varepsilon}}{\sqrt{\tau(1+\varepsilon)}}
  \frac{G_E^p(\tau)}{G_M^p(\tau)}\sin{\theta_2^*}\cos{\phi_2^*}
  +\cos{\theta_2^*} }
  \label{eq:pol-b}
\end{eqnarray}
When, as is typically done, the choice is made to employ parallel
($\parallel \,$: $\theta_2^*= 0$) and perpendicular ($\bot \,$:
$\theta_1^*=\pi/2$, $\phi_1^*=0$) kinematics, this provides a way to
determine the ratio of the form factors:
\begin{equation}
 \sqrt{\frac{\tau(1+\varepsilon)}{2\varepsilon}}
 \cdot\frac{A_\bot}{A_\parallel}=\frac{{G_E^p}(\tau)}{{G_M^p}(\tau)}
 \label{eq:asyrat}
\end{equation}
Similar expressions occur when measuring the recoil polarization (see,
for example,~\citep{Arnold:1980zj, Donnelly:1985ry}).

Analogous studies whose goal is to extract the form factors of the
neutron must generally be undertaken by electron scattering from
few-body nuclei. In particular, inclusive quasi-elastic scattering of
polarized electrons from polarized $^3$He, namely,
$^3\vec{He}(\vec{e}, e^\prime)X$, and semi-inclusive quasi-elastic
scattering of polarized electrons from either polarized deuterons or
$^3$He, namely, $^2\vec{H}(\vec{e}, e^\prime n)p$ and
$^3\vec{He}(\vec{e}, e^\prime n)X$, respectively, or with polarization
transfer to final-state neutrons, $^2H(\vec{e} ,e^\prime\vec{n} )p$,
have all been used to provide effectively elastic electron scattering
from neutrons, {\it i.e.,} $\vec{e}+\vec{n}\to e^\prime+n$ and
$\vec{e}+n\to e^\prime+\vec{n}$. Naturally, in these cases some
corrections for nuclear physics effects must be made.  The separation
of the neutron electromagnetic form factors benefits in two ways from
the use of polarized data.  Not only is the sensitivity to two-photon
corrections decreased, but also some of the nuclear model dependence
cancels in the form factor ratio.

Note that, since the form factors occur as interferences in
Eq.~(\ref{eq:polpiece}) and therefore one is not at high $Q^2$
comparing a very small contribution ($G_E^2$) with a very large
contribution ($G_M^2$) as occurs in the Rosenbluth cross section, it
is believed that one is not as sensitive to higher-order corrections
beyond the one-photon-exchange approximation. This is borne out in
modeling of the two-photon effects~\citep{Guichon:2003qm,
Blunden:2005ew, Carlson:2007sp} which indicate that the Rosenbluth
cross section is problematical in this regard, as mentioned above, but
that these corrections are relatively much less important for the
extraction of the form factor ratio using polarization observables and
that, accordingly, using polarization degrees of freedom in elastic ep
scattering can provide a clean separation of the form factors.  Again,
to make this clear, let us use the simple estimate as above.  The
result in Eq.~(\ref{eq:polpiece}) is proportional to
\begin{equation}
  \vec{p}_T\cdot\vec{A}(\tau, \varepsilon; \theta^*, \phi^*)\sim
  \sqrt{\frac{2\varepsilon}{1+\varepsilon}} \xi_p
  \sin{\theta^*}\cos{\phi^*}+ \cos{\theta^*}
  \label{eq:polpieceX}
\end{equation}
and thus, even at $Q^2=5$ (GeV/c)$^2$ where $\xi_p$ was seen to be
about 0.1, the first term (for $\varepsilon$ not too small) is
typically 10\% of the second and higher-order ${\cal O}(\alpha)$ QED
corrections probably make less of an impact on the extraction of the
form factor ratio.

On the theoretical side, exact {\it ab initio} QCD calculations of
$G^{p,n}_{E,M}$ using lattice techniques will eventually be
possible. However, despite the fact that very encouraging results have
been obtained in recent work~\citep{Schroers:2009zc}, a fully
quantitative understanding of the entire set of form factors is
lacking at present. Given this, alternative approaches are typically
taken. For example, light-front methods, quark descriptions and chiral
invariance have been employed by Miller to obtain qualitative
relations and semi-quantitative descriptions of various aspects of the
form factors in both momentum and configuration
space~\citep{Miller:2002ig, Matevosyan:2005fz,
Matevosyan:2005ia, Miller:2007kt, Miller:2007uy, Miller:2008jc,
Miller:2008pk, Rinehimer:2009yv}.

In the present work we draw upon results from form factor models
which use as hadronic building blocks vector mesons together with
coupling to the $\pi \pi$, $\pi \pi \pi$, and $K\bar K$ continua as
given by dispersion relation calculations --- the so-called Vector
Meson Dominance plus Dispersion Relation based models
(VMD+DR)~\citep{Iachello:1972nu, Hohler:1976ax, Gari:1986rj,
Dziembowski:1997vh, Lomon:2001ga, Miller:2002ig, Lomon:2002jx,
Pasquini:2006ib, Lomon:2006sf}.  The most recent versions of these
models have been quite successful in representing the momentum-space
content in the form factors, {\it i.e.,} the behaviors of the form
factors as functions of 4-momentum transfer squared, especially the
models which also incorporate ingredients that provide the correct
asymptotic behavior as $Q^2 \to \infty$ (see Sect.~\ref{sec:VMD}).
For instance, as discussed in more detail later, one sees that, in
some cases, cancellations of various vector meson contributions can
lead to a dipole-like $Q^2$-dependence, which is in good agreement
with the nucleon's magnetic form factors for $Q^2 < 5$~(GeV/c)$^2$.
The proton's electric form factor is known to fall faster than
dipole and, in fact, even the earliest VMD+DR
models~\citep{Iachello:1972nu, Hohler:1976ax} showed this behavior
although the available data did not. At low $Q^2$ the neutron's
electric form factor has a different form from the proton's, since
the net charge in the neutron is zero; again the polarization data
and VMD+DR approaches yield a $Q^2$-dependence for $G_E^n$ which is
only in rough accord over the current experimental range with the
commonly used dipole type approximation, namely, the Galster
form~\citep{Galster:1971kv}. In the most recent fits, such as
in~\citep{Belushkin:2006qa, Lomon:2006sf} where the high-$Q^2$
behavior predicted by perturbative QCD is enforced, all four of the
nucleon's electromagnetic form factors are very well represented,
showing the experimentally indicated deviations from the dipole or
Galster forms. This will be discussed in more detail in
Sect.~\ref{sec:VMD}. Additionally, a few remarks will be made there
concerning the differences between the VMD+DR approach with hadronic
form factors used here for comparison with data~\citep{Lomon:2006sf}
and a version without such form factors where instead one adds
effective vector mesons~\citep{Belushkin:2006qa}.

In addition to discussing the form factors in Sect.~\ref{sec:VMD},
both the measured quantities and the VMD+DR modeling, {\it i.e.} the
momentum-space content, we also discuss results in coordinate space
(see Sect.~\ref{sec:coord}) with the goal being to obtain additional
insights both into the various representations of the data (p versus
n, $G_E$ versus $G_M$, isoscalar versus isovector, u-quark versus
d-quark) and into the roles being played by the various ingredients
in the VMD+DR approach (the different vector mesons, the role of the
coupling to the continuum, the nature of terms that yield the
asymptotic behavior).

The paper is organized in the following way: following this
introduction, in Sect.~\ref{sec:VMD} the reference model is
discussed in some detail. The basic formalism is summarized,
together with a brief discussion of the data-fitting procedure.
Results from the reference model, denoted GKex, are presented in
Sect.~\ref{sec:momspace},
followed by a brief
discussion where the GKex reference model is compared with another
recent model denoted BHM. In Sec~\ref{subsec:insights} the reference
model is used to attempt to gain some insights into how the various
contributions work with or against each other to produce the
observed form factors. The Breit-frame Fourier transforms of
$G_E^{p,n}$ are discussed in Sect.~\ref{sec:coord}, beginning with
some general caveats on the meaning and relevance of representing
results in coordinate space and proceeding in Sect.~\ref{subsec:FT}
to discuss the procedures used to obtain the Fourier transfers
starting with data in momentum space and to estimate the
uncertainties on the resulting coordinate-space representations. In
Sect.~\ref{subsec:coordresults} the resulting Breit-frame densities
are presented and discussed, and alternative representations are
given (isoscalar/isovector, u-quark/d-quark). Again in this section
the reference model is employed to help in understanding how the
various ingredients enter in producing the Breit-frame Fourier
transforms. Finally, in Sect.~\ref{sec:concl} conclusions resulting
from this study are summarized.


\section{The GKex Model}
\label{sec:VMD}

Given the brief introductory discussions in Sect.~\ref{sec:intro} to
place the general problem in context, let us now summarize the
ingredients in the basic model employed in the present work. We
consider only the VMD+DR approach, as this provides a reasonably
successful representation of the nucleon's electromagnetic form
factors. We start by summarizing some of the basic formulas needed
in the discussions to follow. In particular, the electromagnetic
form factors of a nucleon are defined via the expression for the
electromagnetic current matrix element
\begin{equation}
 \left\langle {N(p^\prime)\left| {J_\mu  } \right|N(p)} \right\rangle =
 \overline u (p^\prime)\left[ {\gamma _\mu  F_1^N (Q^2 ) + \frac{i}{{2m_N
 }}\sigma _{\mu \nu } q^\nu  F_2^N } \right]u(p)\, ,
 \label{eq:curr}
\end{equation}
where $q_\mu\equiv p_\mu-p^\prime_\mu$, $Q^2\equiv-q^2_\mu\ge0$ (in
the space-like regime) is the square of the invariant momentum
transfer, $N$ is the neutron, $n$, or proton, $p$, and $F_1^N(Q^2)$
and $F_2^N(Q^2)$ are respectively the Dirac and Pauli form factors,
normalized at $Q^2=0$: $F_1^p(0)=1$, $F_1^n(0)=0$, $F_2^p(0)=\kappa_p$
and $F_2^n(0)=\kappa_n$ where $\kappa_p$ and $\kappa_n$ are the
anomalous magnetic moments for the proton and neutron, respectively.
The Sachs form factors, most directly obtained from experiment, are
then
\begin{eqnarray}
 G_E^N (Q^2 ) &=& F_1^N (Q^2 ) - \tau F_2^N (Q^2 ) \label{eq:ge} \\
 G_M^N (Q^2 ) &=& F_1^N (Q^2 ) + F_2^N (Q^2 ) \label{eq:gm}
\end{eqnarray}
and the combinations measured by the polarization experiments are
\begin{equation}
 R_N \equiv G_E^N /(G_M^N /\mu _N )\, . \label{eq:RN}
\end{equation}
The isoscalar and isovector form factors are, respectively
\begin{eqnarray}
 F_{1,2}^{(0)} (Q^2 ) &\equiv& F_{1,2}^p (Q^2 ) + F_{1,2}^n (Q^2 )
 \label{eq:f0} \\
 F_{1,2}^{(1)} (Q^2 ) &\equiv& F_{1,2}^p (Q^2 ) - F_{1,2}^n (Q^2 )\, .
 \label{eq:f1}
\end{eqnarray}

Electrons couple through photons to the electromagnetic currents
provided by the hadron and quark distributions within the nucleons,
yielding the form factors introduced above.  Because the photon is a
vector particle, at any parity-conserving vertex where it couples with
hadrons it must connect to these hadrons with unit total angular
momentum and negative parity. The photon does not conserve isospin and
so these systems of hadrons may be isoscalar or isovector. The
simplest such vertex connects the photon to a single vector meson
($\rho$, $\omega$, $\phi$, \dots).  It can also couple to systems of
two or three pions or $K{\bar K}$ in a $1^-$ state, which in turn may
couple to a $\rho$, $\omega$ or $\phi$-type meson. Since the latter
are resonances of the multi-meson systems, the strength of the
interaction is largest close to the masses of the vector mesons. In
leading order this is the VMD limit of the photon-hadron
interaction~\citep{Sakurai:1967aa} which will be seen to give a good
representation of the data over most of the present range of momentum
transfers (see below).  However, small but significant corrections can
be expected from multi-pion correlations in the continuum, such as
those that give the $\rho$ meson its width. These contributions can be
calculated using dispersion relations with input from meson-meson
scattering.  At sufficiently high momentum transfers, as perturbative
Quantum Chromodynamics (pQCD) becomes a better approximation than
effective hadrons, photons coupling to quarks provides a better
description and the models must asymptotically have a pQCD behavior.
This transition is handled in various ways by the models, as discussed
later.

The earliest reasonable fit to the available nucleon form factor data
was a VMD model~\citep{Iachello:1972nu} of Iachello, Jackson and Lande
with $\rho$, $\omega$ and $\phi$ vector meson poles.  They
incorporated a single meson/nuclear vertex form factor for all terms,
using various forms that cut-off at high momentum transfer (but none
decreased as rapidly as pQCD).  The width of the rho-meson was
included by modifying the pole term with a form suggested by Frazer
and Fulco~\citep{Frazer:1960zzb}.  A more recent paper by Bijker and
Iachello~\citep{Bijker:2004yu} adds an asymptotic term to the
Pauli-isovector current and modifies the hadronic form factor to
include the asymptotic logarithmic Q dependence.  After refitting
parameters to a larger set of data, the neutron form factors are
substantially improved at the expense of a small worsening in the fit
to the proton form factors compared with earlier
fit~\citep{Iachello:1972nu}.

Shortly after~\citep{Iachello:1972nu} H\"{o}hler and
collaborators~\citep{Hohler:1976ax} used dispersion relations to
obtain the contribution of the $\pi\pi$ continuum giving the $\rho$
meson its width, which they fitted with a simple function of the mass
(Eq.~(4.2) of that reference). The $\omega$ and $\phi$ mesons and
several phenomenological vector mesons were represented by simple
poles. They did not introduce form factors at the strong vertices.
Instead the phenomenological constants (pole masses and residues) were
restricted by conditions of super-convergent behavior at asymptotic
momentum transfers in addition to being optimized to fit the
data. This required the addition of unknown vector meson pole terms.

Recently Meissner and collaborators~\citep{Mergell:1995bf,
Belushkin:2006qa} have extended the H\"{o}hler type model by
considering, in addition to the $\pi\pi$ continuum, the
$K\overline{K}$ and $\rho\pi$ continua , which they find are
adequately represented by simple poles.  They also added
phenomenological vector meson poles and a broad phenomenological
contribution to each isovector form factor at higher masses.  As
before, there are no strong vertex form factors and the asymptotic
momentum transfer behavior is obtained by requiring a cancellation
amongst all of the terms to obtain super-convergence in one fit, and
an explicit pQCD behavior in another version.

Gari and Kr\"{u}mpelmann (GK)~\citep{Gari:1992qw} proposed a model in
which VMD at low momentum transfers was replaced by pQCD at high
momentum transfers, using differing convergence rates of hadronic and
quark form factors. [Also Ref.~\citep{Gari:1986rj} from earlier.]
They obtained a good fit to the data then available using only the
$\rho$, $\omega$ and $\phi$ vector meson poles.  The hadronic (quark)
form factors are required by the strong renormalization corrections at
the vector meson/nucleon (quark) vertices.  The $\phi$ meson-nucleon
hadronic form factor has been constructed imposing the Zweig rule
required by the $s\overline{s}$ quark structure of that meson.  The
inclusion of these vertex form factors was crucial in enabling the
evolution with momentum transfer to the pQCD behavior without an
artificial constraint on the relation between the vector meson pole
parameters.  As an added indication of the validity of this approach,
there was no need for adding several phenomenological vector meson
poles at masses in disagreement with available data.

The physical realism of this model was enhanced by
Lomon~\citep{Lomon:2001ga, Lomon:2002jx, Lomon:2006sf}
by incorporating the following modifications:
\begin{itemize}
 \item The width of the $\rho$ meson was included using the dispersion
 calculation of~\citep{Belushkin:2006qa};
 \item The observed $\rho^\prime$ (1.45 GeV)~\citep{Lomon:2001ga} and
 $\omega^\prime$ (1.419 GeV)~\citep{Lomon:2002jx} vector meson poles
 were included;
 \item In~\citep{Lomon:2002jx} and later the quark-nucleon vertex form
 factor uses the quark-nucleon cut-off, instead of the meson-nucleon
 cut-off used by GK.  Also the vector meson-hadron form factors of GK
 (model 1) were used as being more consistent with the helicity flip
 in the Pauli terms.  In both cases the logarithmic dependence is
 determined by $\Lambda_{QCD}$, which is fixed near the value
 determined by high-energy data.
\end{itemize}
These yielded the so-called GKex (Gari-Kr\"{u}mpelmann extended)
models used in the present work.  In particular, we employ the model
given in~\citep{Lomon:2006sf} as the basis for the present studies.
Note that our motivation in the present work is not so much to
elaborate the fitting procedures discussed in~\citep{Lomon:2002jx},
but to take as given that study and use the model discussed there to
gain a deeper understanding of some of the systematics seen in the
data. No attempt is made in the present work to provide new fits to
the data after 2005, since the world database is soon to be extended
--- the form factor representations are frozen, using the one specific
contemporary VMD+DR model denoted GKex~\citep{Lomon:2006sf}.
Specifically, we wish to obtain better insight into why
the $G_M^{p,n}$ form factors are roughly dipole in character, while
$G_E^p$ is not, and falls faster than dipole. We shall see that this
difference in behavior emerges naturally in the context of the
models discussed. Furthermore, the most modern models of the type
employed here are actual hybrids containing hadronic ingredients as
well as terms which have the correct pQCD behaviors when $Q^2$
becomes large. Within these models one can ask where the cross-over
to this asymptotic behavior occurs.

The GKex model of~\citep{Lomon:2002jx, Lomon:2006sf} is summarized in
the following. Specifically, the form factors in that model are given
by:
\begin{eqnarray}
 F_1^{(0)} (Q^2 ) &\equiv& \frac{{g_\omega  }}{{f_\omega  }}f^{em} (m_\omega  ;Q^2 )f_1^{had} (Q^2 ) \nonumber \\
 &&+ \frac{{g_{\omega^\prime} }}{{f_{\omega^\prime} }}f^{em} (m_{\omega^\prime} ;Q^2 )f_1^{had} (Q^2 ) \nonumber \\
 &&+ \frac{{g_\phi  }}{{f_\phi  }}f^{em} (m_\phi  ;Q^2 )f_1^{had,s} (Q^2 ) \nonumber \\
 &&+ \left[ {1 - \frac{{g_\omega  }}{{f_\omega  }} - \frac{{g_{\omega^\prime} }}{{f_{\omega^\prime} }}}
 \right]f_1^{had,pQCD} (Q^2 )
 \label{eq:f10}
\end{eqnarray}
\begin{eqnarray}
 F_2^{(0)} (Q^2 ) &\equiv& \kappa _\omega  \frac{{g_\omega  }}{{f_\omega  }}f^{em} (m_\omega  ;Q^2 )f_2^{had} (Q^2 ) \nonumber \\
 &&+ \kappa _{\omega^\prime} \frac{{g_{\omega^\prime} }}{{f_{\omega^\prime} }}f^{em} (m_{\omega^\prime} ;Q^2 )f_2^{had} (Q^2 ) \nonumber \\
 &&+ \kappa _\phi  \frac{{g_\phi  }}{{f_\phi  }}f^{em} (m_\phi  ;Q^2 )f_2^{had,s} (Q^2 ) \nonumber \\
 &&+ \left[ {\kappa _s  - \kappa _\omega  \frac{{g_\omega  }}{{f_\omega  }} - \kappa _{\omega^\prime}
 \frac{{g_{\omega^\prime} }}{{f_{\omega^\prime} }} - \kappa _\phi  \frac{{g_\phi  }}{{f_\phi  }}}
 \right]f_2^{had,pQCD} (Q^2 )
 \label{eq:f20}
\end{eqnarray}
\begin{eqnarray}
 F_1^{(1)} (Q^2 ) &\equiv& \frac{{g_\rho  }}{{f_\rho  }}f^{em} (m_{\rho _1 } ;Q^2 )f_1^{had} (Q^2 )
 \left[ {(1 - \alpha _1 ) + \frac{{\alpha _1 }}{{\left( {1 + Q^2 /Q_1^2 } \right)^2 }}} \right] \nonumber \\
 &&+ \frac{{g_{\rho^\prime} }}{{f_{\rho^\prime} }}f^{em} (m_{\rho^\prime} ;Q^2 )f_1^{had} (Q^2 ) \nonumber \\
 &&+ \left[ {1  - \frac{{g_\rho  }}{{f_\rho  }} - \frac{{g_{\rho^\prime} }}{{f_{\rho^\prime} }}}
 \right]f_1^{had,pQCD} (Q^2 )
\label{eq:f11}
\end{eqnarray}
\begin{eqnarray}
 F_2^{(1)} (Q^2 ) &\equiv& \kappa _\rho  \frac{{g_\rho  }}{{f_\rho  }}f^{em} (m_{\rho _2 } ;Q^2 )f_2^{had} (Q^2 )
 \left[ {(1 - \alpha _2 ) + \frac{{\alpha _2 }}{{\left( {1 + Q^2 /Q_2^2 } \right)}}} \right] \nonumber \\
 &&+ \kappa _{\rho^\prime} \frac{{g_{\rho^\prime} }}{{f_{\rho^\prime} }}f^{em} (m_{\rho^\prime} ;Q^2 )f_2^{had} (Q^2 ) \nonumber \\
 &&+ \left[ {\kappa _v - \kappa _\rho  \frac{{g_\rho  }}{{f_\rho  }} - \kappa _{\rho^\prime} \frac{{g_{\rho^\prime} }}{{f_{\rho^\prime} }}}
 \right]f_2^{had,pQCD} (Q^2 )\, .
 \label{eq:f21}
\end{eqnarray}
In these expressions the anomalous magnetic moments are $\kappa _s =
\kappa _p + \kappa _n$ and  $\kappa _v = \kappa _p - \kappa _n$, and
the $\kappa_x$ are the analogous quantities associated with the vector
mesons $x = \rho$, $\rho^\prime$, $\omega$, $\omega^\prime$ and
$\phi$.  The pole corresponding to a vector meson of mass $m_x$ yields
the monopole form
\begin{equation}
 f^{em} (m_x ;Q^2 ) \equiv  \left[ {\frac{{m_x^2 }}{m_x^2  + Q^2 }} \right] \quad x = \rho ,\rho^\prime,\omega
,\omega^\prime,\phi
 \label{eq:vmp}
\end{equation}
and the coupling constant of each pole is $g_{x}/f_{x}$, $x = \rho$,
$\rho^\prime$, $\omega$, $\omega^\prime$ and $\phi$, where $g_{x}$ is
the coupling of meson to the nucleon and $f_{x}$ is given by the
coupling of the meson to the photon.  The value of $f_{x}$ is
experimentally determined from the meson decay to $e^+ e^-$.

For completeness we briefly summarize the procedures used
in~\citep{Lomon:2001ga, Lomon:2002jx, Lomon:2006sf} to determine the
model parameters.  Specifically, the 2001 version of the GKex model,
which did not include the $\omega^\prime$ meson, was fitted to all of
the unpolarized, Rosenbluth-separated cross section data, and included
the then-available $R_p$ polarization data, although in the absence of
$R_n$ data. The 2002 GKex model includes the then-available
polarization $R_p$ and $R_n$ data, some of which was not final. The
present 2005 GKex model --- the one used as a basis for the present
study --- differs from the 2002 version only due to the substitution
of the final polarization data, inclusion of the few new $R_n$ and
$G_M^n$ points, and the exclusion of the higher $Q^2$ $G_E^{p,n}$ data
from the Rosenbluth separation of differential cross section data. For
completeness we list the parameters obtained using the last
model~\citep{Lomon:2006sf}. Given the fact that new data will soon be
available, no re-fitting has been done for the present study, although
it is anticipated that this will be performed in the near future. The
masses of the known vector mesons are fixed: $m_{\rho} = 0.776$ GeV,
$m_{\omega} = 0.784$ GeV, $m_{\rho^\prime} = 1.45$ GeV,
$m_{\omega^\prime} = 1.419$ GeV and $m_{\phi} = 1.019$ GeV. The ratios
$g/f$ are as follows: $g_{\rho} / f_{\rho} = 0.5596$, $g_{\omega} /
f_{\omega} = 0.7021$, $g_{\rho^\prime} / f_{\rho^\prime} = 0.0072089$,
$g_{\omega^\prime} / f_{\omega^\prime} = 0.164$ and $g_{\phi} /
f_{\phi} = -0.1711$.  The vector mesons' anomalous magnetic moments
are $\kappa_{\rho} = 5.51564$, $\kappa_{\omega} = 0.4027$,
$\kappa_{\rho^\prime} = 12.0$, $\kappa_{\omega^\prime} = -2.973$,
$\kappa_{\phi} = 0.01$ and one finds that $\mu_{\phi} = 0.2.$

Defining
\begin{equation}
 {\widetilde Q}^2  \equiv Q^2 \frac{{\ln \left[ {\left( {\Lambda_D^2
 + Q^2 } \right)/\Lambda _{QCD}^2 } \right]}}{{\ln \left[ {\Lambda_D^2 /\Lambda _{QCD}^2 } \right]}}\, , \label{eq:qtilde}
\end{equation}
with $\Lambda_D=1.181$~GeV and $\Lambda_{QCD}=0.150$~GeV (fixed),
thereby incorporating the logarithmic momentum transfer behavior of
pQCD, the hadronic vector-meson to nucleon form factors for those
vector mesons dominantly consisting of non-strange quarks ($\rho$,
$\omega$, $\rho^\prime$ and $\omega^\prime$) are given by
\begin{eqnarray}
 f_1^{had} (Q^2 ) &\equiv& f(\Lambda _1 ;{\widetilde Q}^2 )f(\Lambda _2 ;{\widetilde Q}^2 ) \label{eq:fhad1}\\
 f_2^{had} (Q^2 ) &\equiv& f(\Lambda _1 ;{\widetilde Q}^2 )f(\Lambda _2 ;{\widetilde Q}^2 )^2\, ,
 \label{eq:fhad2}
\end{eqnarray}
where
\begin{equation}
f(\Lambda _i ;{\widetilde Q}^2 ) \equiv  \left[ {\frac{{\Lambda_i^2 }}{\Lambda_i^2  + {\widetilde Q}^2 }} \right]\, ,
 \label{eq:ns}
\end{equation}
{\it i.e.} functionally the same (monopole) expression as
Eq.~(\ref{eq:vmp}), now with $m_x \to \Lambda_i$ and $Q^2 \to
{\widetilde Q}^2$. From the fit one has $\Lambda_1=0.93088$~GeV and
$\Lambda_2=2.6115$~GeV.\footnote{The constants used in the GKex
model are given here to high precision not because they are so well
known, but because they will allow others to program the formulas in
this section and check their results against the results found in the
present study.} The spin-flip nature of the Pauli term in the current
is the origin of the extra power of $f(\Lambda _2 ;{\widetilde Q}^2 )$
in Eq.~(\ref{eq:fhad2}).

For the $\phi$ meson, which is dominantly composed of strange quarks,
the hadronic form factors are given by
\begin{eqnarray}
 f_1^{had,s} (Q^2 ) &\equiv& f_1^{had} (Q^2 )\left[ {\frac{{Q^2 }}{{\Lambda _1^2  + Q^2 }}} \right]^{3/2}
  \label{eq:fhads1}\\
 f_2^{had,s} (Q^2 ) &\equiv& f_2^{had} (Q^2 )\left[ {\frac{{\mu _\phi ^2  + Q^2 }}{{\mu _\phi ^2 }}
 \frac{{\Lambda _1^2 }}{{\Lambda _1^2  + Q^2 }}} \right]^{3/2}\, .
 \label{eq:fhads2}
\end{eqnarray}
The form factor $f_1^{had,s}$ vanishes at $Q^2 = 0$, and it and
$f_2^{had,s}$ decrease more rapidly at large $Q^2$ than the other
meson form factors. This conforms to the Zweig rule imposed by the
$s\overline{s}$ structure of the $\phi$
meson~\citep{Christy:2004rc}. Only 10 of the 12 parameters listed
above are independent, as $\kappa_\phi / \mu_\phi$ and
$\kappa_{\rho^\prime} g_{\rho^\prime} / f_{\rho^\prime}$ are
constrained to be very close to 0.05 and 0.08, respectively. The fit
has little sensitivity to $\Lambda_{QCD}$, which is fixed at 0.150 in
its experimental range.

All of the terms but two in the above isoscalar and isovector form
factors are of the pole form representing a vector meson exchange.
However, the first term in each of the isovector form factors is an
approximate analytic form for a $\rho$ meson with a width derived from
a dispersion integral of the $\pi\pi$ continuum.  For later
discussions, we have written these expressions using parameters
$\alpha_1$ ($\alpha_2$) for the $F_1^{(1)}$ ($F_2^{(1)}$) expressions,
respectively, where $\alpha_1 =0.0781808$ and $\alpha_2 =0.0632907$
when the widths are included, with $\alpha_i =0$, $i=1,$ 2 when the
effect from coupling to the continuum is ignored. In addition, when
the contributions from the continuum are included, the effective
$\rho$ mass is shifted down slightly from the physical mass:
$m_{\rho_i}=m_{\rho} - \delta_i$ with $\delta_1 = 34.65$ MeV and
$\delta_2 = 43.74$ MeV. When the $\rho$ contributions are taken to
occur only at the pole, of course these shifts are also neglected and
the physical mass used in the expressions above. The momentum cutoffs
in the terms that occur when the width is included are $Q_1^2= 0.3176$
(GeV/c)$^2$ and $Q_2^2= 0.1422$ (GeV/c)$^2$. All of these constants
are determined by a dispersion calculation and we use the results
obtained by~\citep{Mergell:1995bf}. Note that turning off the width
and using only the $\rho$-pole form is not fully consistent: one
should refit the data with the $\alpha_i =0$ to do this correctly.
However, for our present purposes simply turning the width off gives
us some indication of where one might expect the coupling to the
$\pi\pi$ continuum to play a role, either in momentum space or in
coordinate space.

For the asymptotic terms, the form factors due to the coupling of the
mesons to the nucleons at the quark level are given by
\begin{eqnarray}
 f_1^{had,pQCD} (Q^2 ) &\equiv& f(\Lambda _D ;{\widetilde Q}^2 )f(\Lambda _2 ;{\widetilde Q}^2 ) \label{eq:fhadpqcd1}\\
 f_2^{had,pQCD} (Q^2 ) &\equiv& f(\Lambda _D ;{\widetilde Q}^2 )f(\Lambda _2 ;{\widetilde Q}^2 )^2\, .
 \label{eq:fhadpqcd2}
\end{eqnarray}
The coefficients of these terms impose the constraints at $Q^2 =0$,
\begin{eqnarray}
 F_1^{(0)} (0) &=& F_1^{(1)} (0) = 1 \nonumber \\
 F_2^{(0)} (0) &=& \kappa _s \qquad\qquad F_2^{(1)} (0) = \kappa _v\, ,
 \label{eq:ffatzero}
\end{eqnarray}
and when $Q^2 \rightarrow \infty$ have the asymptotic forms
\begin{eqnarray}
 F_1^{(0,1)} \rightarrow \frac{1}{{Q^2 \ln \left( {Q^2 /\Lambda _{QCD}^2 } \right)}} \nonumber \\
 F_2^{(0,1)} \rightarrow \frac{1}{{Q^4 \ln \left( {Q^2 /\Lambda _{QCD}^2 } \right)}}
 \label{eq:ffatinfty}
\end{eqnarray}
as required by pQCD.

The GKex model employed in the present study is the one
of~\citep{Lomon:2006sf} with the parameters fitted to a large data
set, for which the low-$Q^2$ BLAST data were not yet available.
Included in the data set were $G^p_M$ and $G^n_M$ from Rosenbluth
separations of unpolarized cross sections, and $R_p$ and $R_n$
obtained from polarization measurements, over the whole experimental
energy range. The $G^p_E$ and $G^n_E$ results obtained by Rosenbluth
separation of the unpolarized cross sections were only included at
lower $Q^2$ where they are more than a few percent of the magnetic
cross section, and therefore not too sensitive to the two-photon
contributions discussed in Sect.~\ref{sec:intro}. At higher $Q^2$
the $G^{p,n}_E$ from the Rosenbluth separations are systematically
larger than those obtained by multiplying the polarization
observables, $R_{p,n}$ by the $G^{p,n}_M$ obtained from the
unpolarized cross sections. A recent higher accuracy
measurement~\citep{Qattan:2004ht} of the unpolarized cross section
confirms this result.

In detail, the data from Refs.~7--14, and~16--36 cited
in~\citep{Lomon:2001ga} were used, with the omission of the $G_E^p$
values for $Q^2\geq 1.75$ (GeV/c)$^2$ of Ref.~7 and the $G_E^n$ values
for $Q^2\geq 0.779$ (GeV/c)$^2$ of Refs~9, 17, and~18 there.
Reference~\citep{Lomon:2006sf} used the $R_p$ values of Ref.~5, the
$R_n$ values of Refs.~4 and~6 and the recent $G_M^n$ data of Ref.~7.
It should be emphasized that the form factor data sets were all fit
simultaneously. Another datum used is the slope $dG_E^n/dQ^2 (Q^2 =
0) =$ 0.0199 $\pm$ 0.0003~fm$^2$, as determined by thermal neutron
scattering~\citep{Kopecky:1995zz, Aleksandrov:1999jz}. Although this
is the most accurate $G_E^n$ information, it is often not considered
in model fitting.


\section{Results in Momentum Space and Comparisons with Data}
\label{sec:momspace}

\begin{figure}[hbt]
  \includegraphics[width=\textwidth]{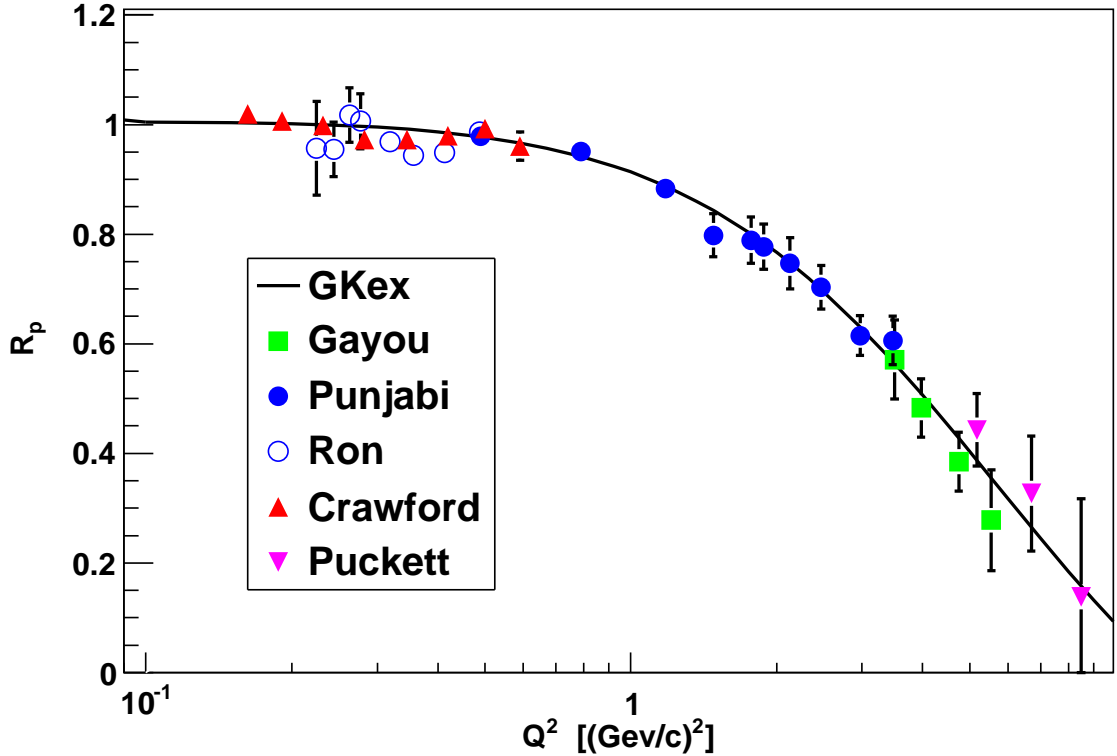}
  \caption{Form factor ratio $R_p$ showing the GKex universal
  fit~\citep{Lomon:2006sf} together with the fitted data
  (Gayou~\citep{Gayou:2001qd}, Punjabi~\citep{Punjabi:2005wq}, and
  Ron~\citep{Ron:2007vr}; see the end of Sect:~\ref{sec:VMD}) and
  recent measurements from BLAST (Crawford~\citep{Crawford:2006rz})
  and JLab (Puckett~\citep{Puckett:2010ac}). In color online.}
  \label{elFig1}
\end{figure}

Fig.~\ref{elFig1}
shows $R_p$ as represented by the GKex model~\citep{Lomon:2006sf}
(fitted to the data listed at the end of
Sect.~\ref{sec:VMD}) together with the polarization
data~\citep{Gayou:2001qd,Punjabi:2005wq,Crawford:2006rz,Ron:2007vr}.
The $R_p$ data used in the fit were the polarization measurements
of~\citep{Gayou:2001qd, Punjabi:2005wq} and (not shown) the ratio
extracted from a Rosenbluth separation~\citep{Bartel:1973rf}, while
the results presented in~\cite{Crawford:2006rz, Ron:2007vr} were not
used in the fit. The model fits the polarization data well while not
conforming to the results obtained from Rosenbluth separations.
Moreover, as shown, this fit predicted the new BLAST low momentum
transfer results~\citep{Crawford:2006rz, Ron:2007vr} well and is in
excellent agreement with the very recent
results at higher $Q^2$ from JLab~\citep{Puckett:2010aa}. The
deviation from unity is substantial for $Q^2>0.8$ (GeV/c)$^2$;
indeed, as stated in the previous section, this has always been a
feature of the VMD class of models in that from their inception they
have typically led to a fall-off with $Q^2$ of $G_E^p$ compared with
the dipole form factor.

\begin{figure}[htbp]
 \includegraphics[width=\textwidth]{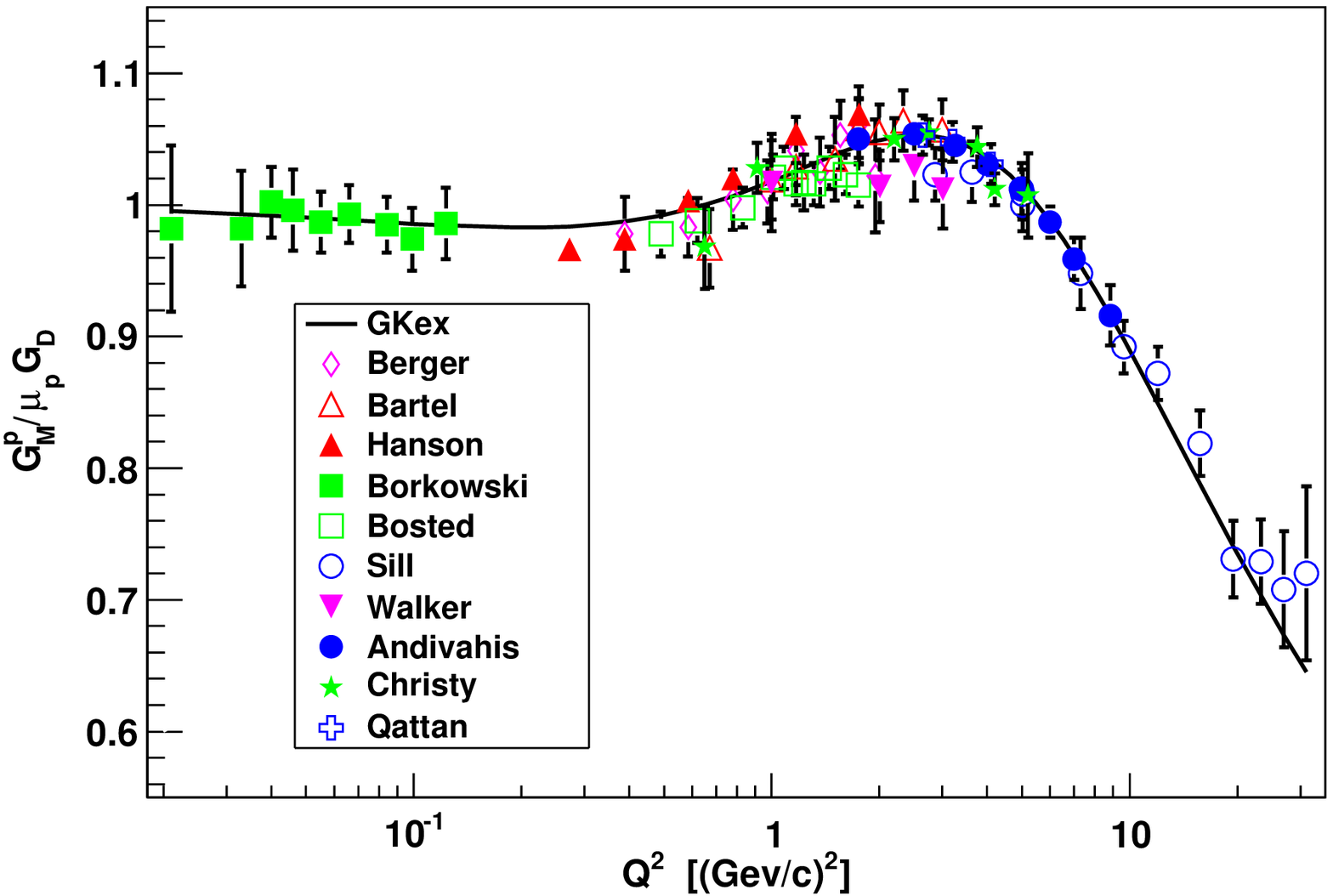}
 \caption{The ratio $G_M^p/\mu_p G_D$ with the GKex universal
 fit~\citep{Lomon:2006sf} and the fitted data
 (Berger~\citep{Berger:1971kr}, Bartel~\citep{Bartel:1973rf},
 Hanson~\citep{Hanson:1973vf}, Borkowski~\citep{Borkowski:1974mb},
 Bosted~\citep{Bosted:1989hy}, Sill~\citep{Sill:1992qw},
 Walker~\citep{Walker:1993vj}, and Andivahis~\citep{Andivahis:1994rq}; see the end of Sect:~\ref{sec:VMD}).
 In addition the data from Christy~\citep{Christy:2004rc} and the more
 recent precision data Qattan~\citep{Qattan:2004ht} are shown. In
 color online.}
\label{elFig2}
\end{figure}

Fig.~\ref{elFig2}
displays the model result for $G_M^p/\mu_p G_D$, where $G_D$ is the
standard dipole form.  The model was fitted to all the Rosenbluth
determinations of $G_M^p$ data~\citep{Berger:1971kr, Bartel:1973rf,
Hanson:1973vf, Borkowski:1974mb, Bosted:1989hy, Sill:1992qw,
Walker:1993vj, Andivahis:1994rq}.  In addition the data
from~\citep{Christy:2004rc} and the more recent precision
data~\citep{Qattan:2004ht} is shown.  The momentum transfer range is
greater than for the other form factors.  The ratio is relatively
close to unity until $Q^2\approx1$ when it increases before decreasing
rapidly for $Q^2>7$~(GeV/c)$^2$.

\begin{figure}[htbp]
 \includegraphics[width=\textwidth]{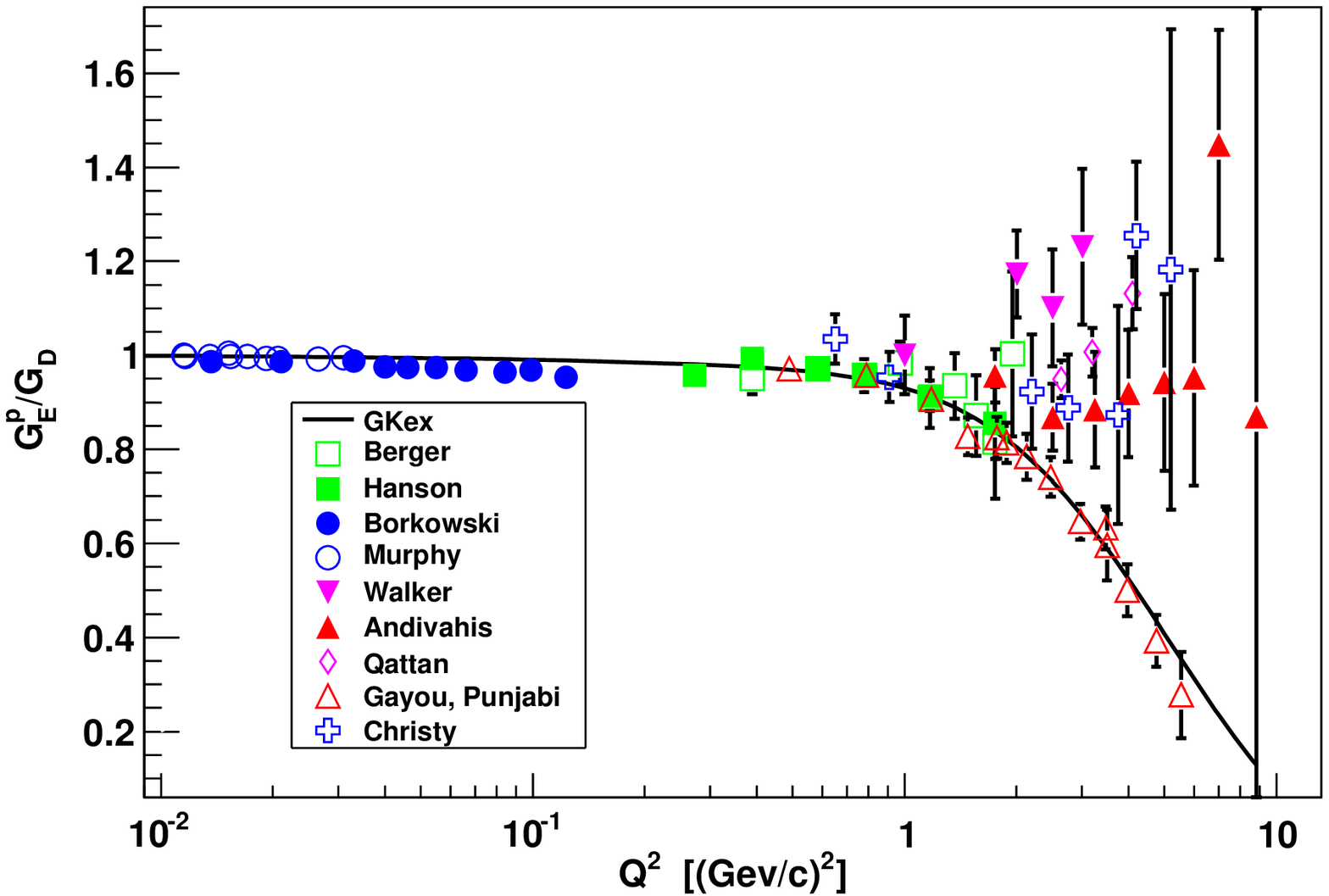}
 \caption{The ratio $G_E^p/G_D$ with the GKex universal
 fit~\citep{Lomon:2006sf}.  Because the higher momentum-transfer
 values obtained from the Rosenbluth separation are not considered
 reliable, only the lower momentum-transfer values from Rosenbluth
 separation (Berger~\citep{Berger:1971kr},
 Hanson~\citep{Hanson:1973vf}, Borkowski~\citep{Borkowski:1974mb}, and
 Murphy~\citep{Murphy:1974zz}) were included in the fitting. However,
 the higher momentum-transfer values (Walker~\citep{Walker:1993vj},
 Andivahis~\citep{Andivahis:1994rq}, and Qattan~\citep{Qattan:2004ht}; see the end of Sect:~\ref{sec:VMD})
 are also plotted.  The fitted $R_p$ data of Gayou~\citep{Gayou:2001qd}
 and Punjabi~\citep{Punjabi:2005wq} were translated to $G_E^p$ by
 multiplying by the GKex $G_M^p/\mu_p$. Data from
 Christy~\citep{Christy:2004rc} is also shown. In color online.}
 \label{elFig3}
\end{figure}

Fig.~\ref{elFig3}
shows the model results for $G_E^p/G_D$.  The model was fitted to
the low-$Q^2$ $G_E^p$ differential cross section data
of~\citep{Berger:1971kr, Hanson:1973vf, Borkowski:1974mb,
Murphy:1974zz}.  For the reasons given above (small contribution
to the unpolarized cross section and 2-photon corrections) the
higher-$Q^2$ data displayed~\citep{Walker:1993vj,
Andivahis:1994rq, Qattan:2004ht} were not included in the fitting
procedure.  Also shown are data~\citep{Christy:2004rc} and the
$G_E^p$ values given by the polarization values of
$R_p$~\citep{Gayou:2001qd, Punjabi:2005wq} multiplied by the model
$G_M^p/\mu_p$. Above 1.8 (GeV/c)$^2$ the model fits the
polarization values, but not those obtained from Rosenbluth
separations.

The extraction of the neutron form factors from quasi-elastic
electron-deuteron or electron-$^3$He scattering, with their dependence
on the nuclear wave function and hadronic final-state interactions,
leads to greater uncertainties and a more restricted momentum transfer
range than for the proton form factor. There is also some evidence at
the highest available momentum transfers of the deviation from the
dipole form for the magnetic form factor, and from the modified dipole
(Galster) form for the electric-to-magnetic ratio.

\begin{figure}[htbp]
  \includegraphics[width=\textwidth]{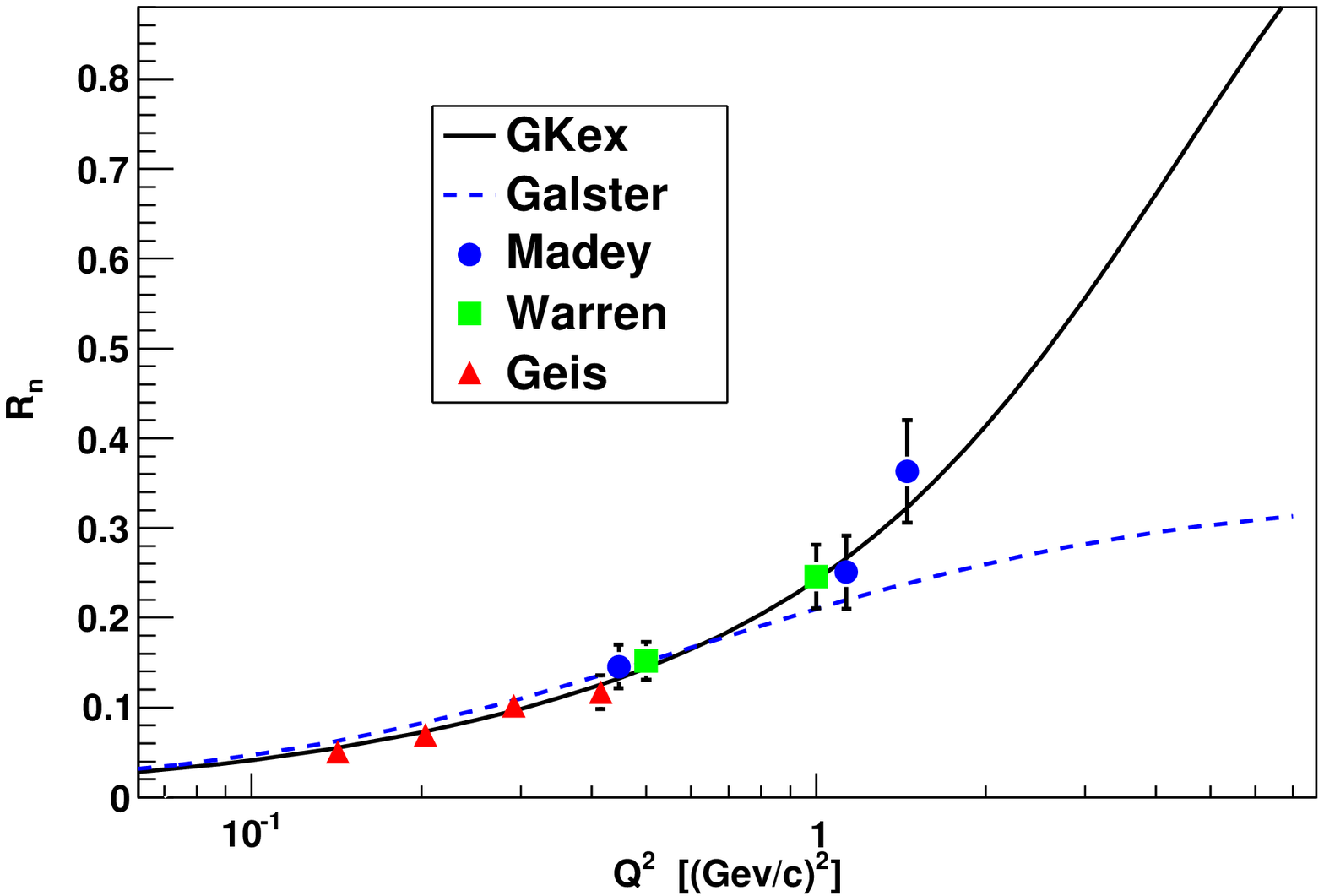}
  \caption{Form factor ratio $R_n$ with the GKex universal
  fit~\citep{Lomon:2006sf} and the Galster~\citep{Galster:1971kv}
  parameterization with the fitted data (Madey~\citep{Madey:2003av}
  and Warren~\citep{Warren:2003ma}; see the end of Sect:~\ref{sec:VMD}) and recent results from BLAST
  (Geis~\citep{Geis:2008ha}). In color online.}
  \label{elFig4}
\end{figure}

Fig.~\ref{elFig4}
shows $R_n$ given by the GKex model~\citep{Lomon:2006sf}.  In that
model only the polarization data of~\citep{Madey:2003av,Warren:2003ma}
were fitted, but not the more recent low-$Q^2$ BLAST
data~\citep{Geis:2008ha} nor the preliminary higher-$Q^2$ JLab
data~\citep{Wojtsekhowski:2008aa}.  Nevertheless, the 2005 fit agrees
very well with the BLAST results and with the preliminary data (not
shown). The Galster form (dashed curve) is also shown, the slope of
which at $Q^2=0$ is known to be larger than that obtained from cold
neutron scattering.  As seen in the figure this results in the Galster
curve being above the BLAST data and the model curve up to
0.4~(GeV/c)$^2$ .  Above that momentum transfer the Galster expression
drops below the data and the model curve.

\begin{figure}[htbp]
 \includegraphics[width=\textwidth]{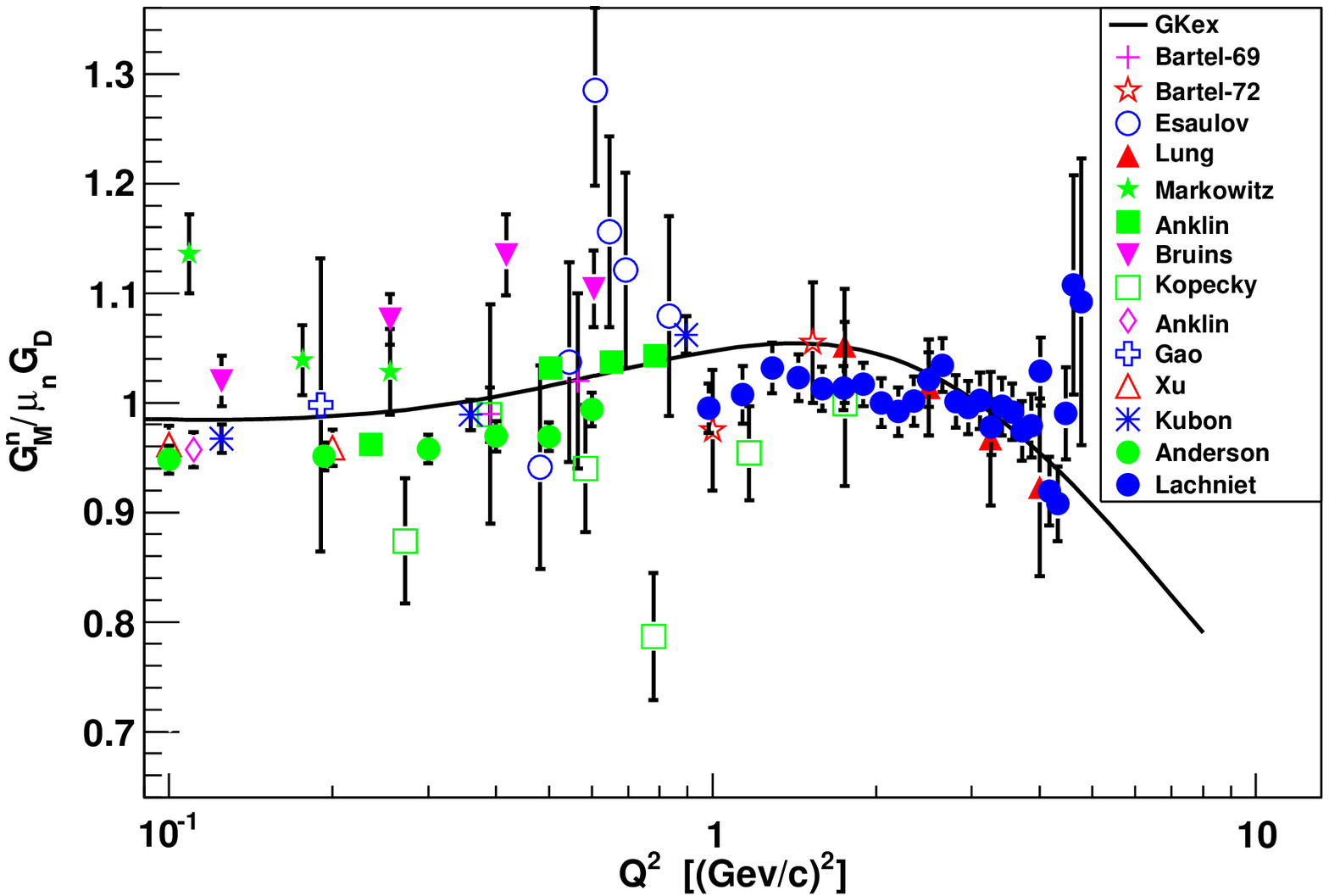}
 \caption{The ratio $G_M^n/\mu_n G_D$ with the GKex universal
 fit~\citep{Lomon:2006sf} and the fitted data
 (Bartel-69~\citep{Bartel:1969an}, Bartel-72~\citep{Bartel:1972xq},
 Esaulov~\citep{Esaulov:1987uc}, Lung~\citep{Lung:1992bu},
 Markowitz~\citep{Markowitz:1993hx}, Anklin~\citep{Anklin:1994ae},
 Bruins~\citep{Bruins:1995ns}, Kopecky~\citep{Kopecky:1995zz},
 Anklin~\citep{Anklin:1998ae}, Gao~\citep{Gao:1998in},
 Xu~\citep{Xu:2000xw}, Kubon~\citep{Kubon:2001rj}; see the end of Sect:~\ref{sec:VMD}) together with new
 results from JLab (Anderson~\citep{Anderson:2006jp} and
 Lachinet~\citep{Lachniet:2008qf}). In color online.}
 \label{elFig5}
\end{figure}

All of the $G_M^n$ data~\citep{Bartel:1969an, Bartel:1972xq,
Esaulov:1987uc, Lung:1992bu, Markowitz:1993hx, Anklin:1994ae,
Bruins:1995ns, Kopecky:1995zz, Anklin:1998ae, Gao:1998in, Xu:2000xw,
Kubon:2001rj}, except the recent JLab data~\citep{Anderson:2006jp,
Lachniet:2008qf}, were used in the 2005 fit.  As seen in
Fig.~\ref{elFig5},
below 1~(GeV/c)$^2$ the data are inconsistently scattered even
within individual data sets.  The model tracks an average of the
scattered data and fits the higher-$Q^2$ data well, dropping below
the dipole values above $Q^2=4$~(GeV/c)$^2$ .  The newer
data~\citep{Lachniet:2008qf} are a little lower in the mid-range
and this reinforces the tendency to go below the dipole fit.

\begin{figure}[htbp]
  \includegraphics[width=\textwidth]{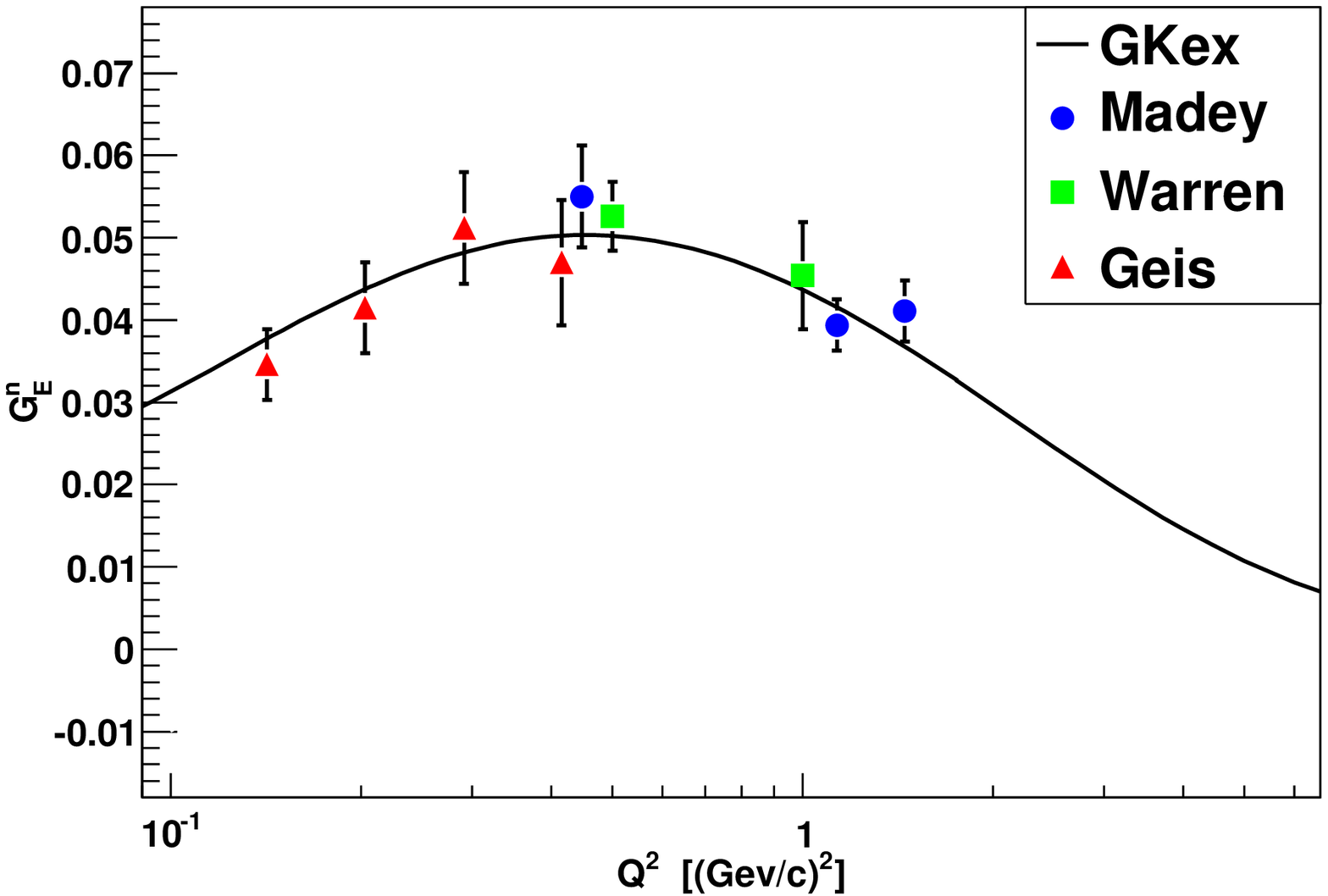}
  \caption{$G_E^n$ with the GKex universal fit~\citep{Lomon:2006sf}.
  The inconsistent $G_E^n$ data from the unpolarized differential
  cross section fitted in~\citep{Lomon:2006sf} are not plotted here.
  The data points are translated from Fig.~\ref{elFig4} through
  multiplication of $R_n$ by $G_M^n/\mu_n$. In color online.}
  \label{elFig6}
\end{figure}

Fig.~\ref{elFig6}
shows that $G_E^n$ , just as $G_E^p/G_D$ in Fig.~\ref{elFig3} fits the
data derived from polarization results of Fig.~\ref{elFig4} very well.
The values obtained from Rosenbluth separations~\citep{Bartel:1969an,
Bartel:1972xq, Hanson:1973vf, JonesWoodward:1991ih, Lung:1992bu,
Eden:1994ji, Passchier:1999cj, Rohe:1999sh, Becker:1999tw,
Herberg:1999ud, Ostrick:1999xa, Golak:2000nt, Zhu:2001md} would be
much higher than those, but are not plotted because of their greater
sensitivity to the two-photon corrections and the nuclear target model
dependence.

Figures~\ref{elFig1}--\ref{elFig6} show not only the data at
low-$Q^2$, the main focus of this study, but also over an expanded
range to see the small structures in the data and models better.  It
is noteworthy that, while the parameters of this model were fitted
using the whole momentum transfer region of the available data, the
model reproduces the low momentum transfer BLAST data recently
obtained (after the model fit) for $Q^2$ between~0.1 and~0.6
(GeV/c)$^2$~\citep{Crawford:2006rz, Geis:2008ha}.  These new data do
not confirm possible ``bump" structures near 0.2 (GeV/c)$^2$ suggested
by earlier measurements and the invocation of a phenomenological pion
cloud~\citep{Friedrich:2003iz} is not required.  In VMD-DR models,
such as the ones discussed here, the pion cloud is represented by pion
pairs and triplets largely clustered into vector mesons. 
This is consistent with the analysis of Hammer, Drechsel,
Meissner~\citep{Hammer:2003qv} which shows that, after the imposition
of unitarity, the addition of $\pi\pi$ continuum to that given by the
$\rho$ is insufficient to provide a substantial bump structure.

Finally, a few words are in order concerning the full GKex form
factors and their pQCD terms.  Because $\Lambda_{\mathrm{QCD}}$ is
$\leq$ 200~MeV, it was initially expected that the asymptotic pQCD
region would be approached at momentum transfers not much larger
than 1~GeV/c~\citep{Lepage:1979za, Lepage:1980fj}.  This may apply
to inclusive reactions, but it was pointed out~\citep{Isgur:1984jm,
Isgur:1988iw, Isgur:1989cy} that for exclusive processes the
momentum transfer had to be shared among several exchanged gluons.
It was then estimated that pQCD may not be approached for elastic
form factors until the order of 1000~GeV/c. In fact for elastic
proton-proton scattering the strong persistence of polarization
effects~\citep{Crabb:1990as} (which vanish in pQCD) at
$T_{lab}=28$~GeV involves much larger momentum transfers, up to
8~(GeV/c)$^2$.
\begin{figure}[htbp]
  \includegraphics[width=\textwidth]{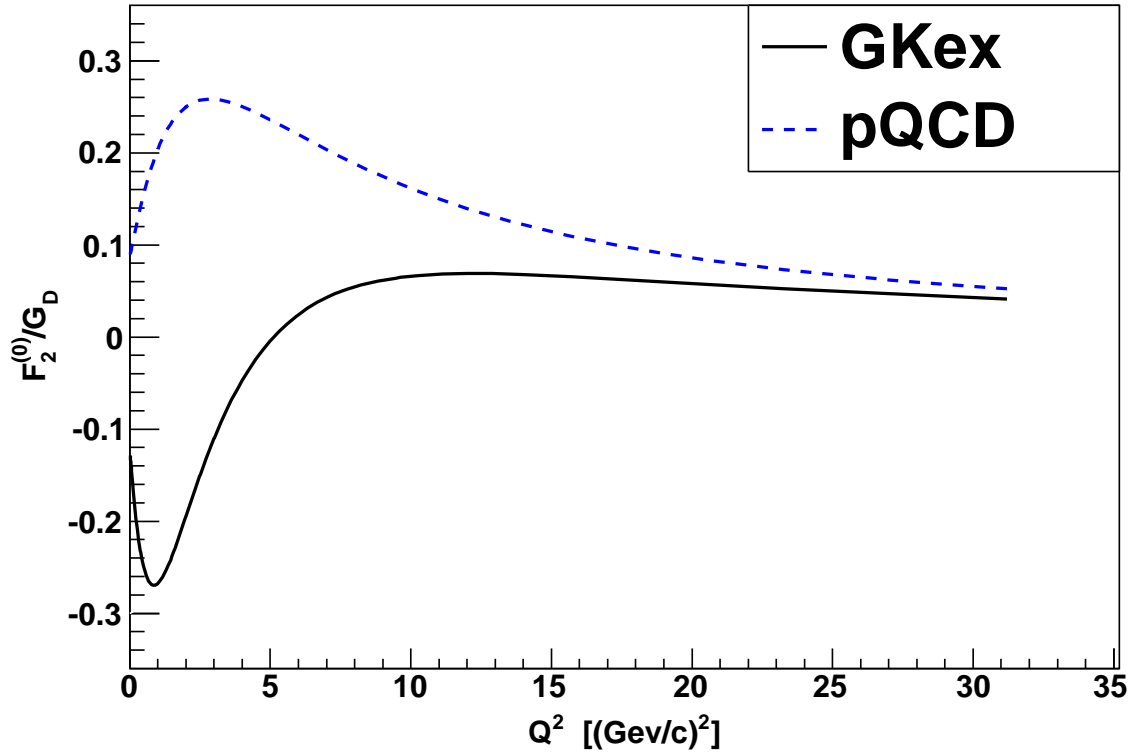}
  \caption{The isoscalar Pauli form factor normalized to $G_D$ with
  the GKex universal fit~\citep{Lomon:2006sf} (solid) and the pQCD
  term of that fit (dashed). In color online.}
  \label{fig:elFig7}
\end{figure}

For this model and its normalization of the pQCD limit, the magnetic
form factors and pQCD are about 10\% different at
$Q^{2}\sim10$~(GeV/c)$^2$. While $R_p$ is within 10\% of pQCD near
2~(GeV/c)$^{2}$, $R_{n}$ is only 80\% of pQCD at 50~(GeV/c)$^{2}$.
Separating the isovector and isoscalar and the Dirac and Fermi terms
gives a more specific indication of the slow approach to pQCD, as
doing so minimizes accidental cancellations between terms.
The isovector form factors $F^{(1)}_{1,2}$ are both
relatively large. One finds that for $Q^2<5$ (GeV/c)$^2$ three of
the four form factors are very different from the pQCD results alone
--- only $F^{(1)}_{1}$ is relatively similar to the pQCD contribution
down to about 2 (GeV/c)$^2$. As $Q^2$ increases beyond about 5 (GeV/c)$^2$ the pQCD
contribution begins to saturate the total; specifically, at 10
(GeV/c)$^2$ the ratio of the pQCD contribution to the total is 96\%
for $F^{(1)}_{1}$ and 83\% for $F^{(1)}_{2}$. The corresponding
numbers at 20 (GeV/c)$^2$ are 98\% and 88\%, respectively. The
isoscalar form factor $F^{(0)}_{1}$ is somewhat smaller than the
isovector form factors and again shows saturation of the pQCD
contribution with increasing $Q^2$, although somewhat more slowly
than for the isovector form factors. The ratio of the pQCD
contribution to the total result for $F^{(0)}_{1}$ is 79\% at 10
(GeV/c)$^2$ and 88\% at 20 (GeV/c)$^2$. Finally, the isoscalar form
factor $F^{(0)}_{2}$ is relatively small and slower to converge to
the pQCD result (see Fig.~\ref{fig:elFig7}). It should also be
noted that the model curve for $F^{(0)}_{2}$ has a substantial dip
near 1~(GeV/c)$^{2}$ which can be attributed to the opposite signs
of the large $\omega$ and $\omega^\prime$ magnetic contributions. In
Sect.~\ref{subsec:insights} we show the individual contributions to
the form factors, including those from the pQCD terms discussed
here. The convergence is similar for the previous GKex
model~\citep{Lomon:2002jx}.  However, the
pQCD normalization is expected to depend on possible major
modifications of the model such as the addition of non-pQCD terms
above the vector meson resonance region.


Finally, recently Belushkin, Hammer and
Meissner~\citep{Belushkin:2006qa} [BHM] have extended the
H\"{o}hler-type model by considering the K\={K} and $3\pi$ continua
in addition to the $2\pi$ continuum, and conclude that the first two
are adequately represented by including only simple poles and adding
a broad phenomenological contribution to each isovector form factor
at higher masses. The asymptotic momentum transfer behavior is
restricted by a super-convergent requirement in one fit, but by an
explicit pQCD behavior in another version. As there are no hadronic
form factors, the required asymptotic behavior is obtained by a
restriction on the sum of all terms in the fit to the coupling
strengths and masses. This results in requiring vector mesons with
unobserved masses.  The BHM-pQCD asymptotic behavior model requires
fewer extra vector mesons than the BHM super-convergent (SC) model.

\begin{figure}[htbp]
 \includegraphics[width=\textwidth]{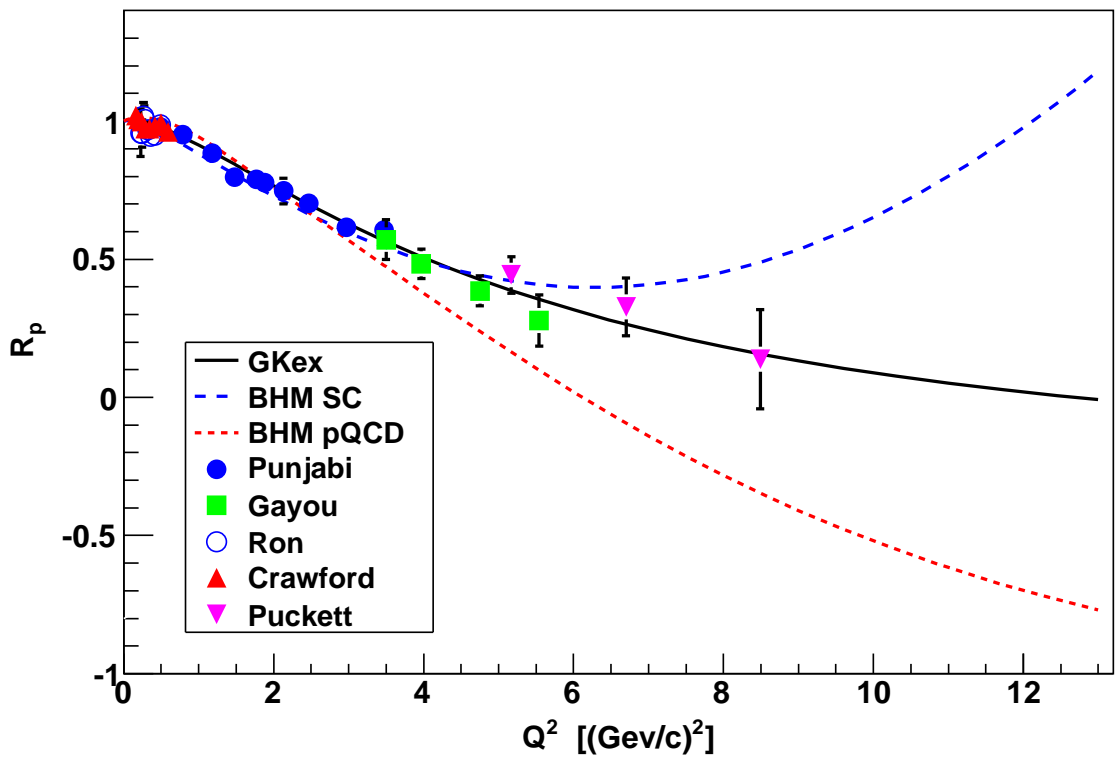}
 \caption{$R_p$ compared to the GKex~\citep{Lomon:2006sf} and
 BHM~\citep{Belushkin:2006qa} super-convergent and asymptotically pQCD
 curves with the data as in Fig.~\ref{elFig1}. In color online.}
  \label{elFig8}
\end{figure}

Overall the GKex model agrees with the data better than
do either the BHM-pQCD or BHM-SC models. Figure~\ref{elFig8}
illustrates the above remarks for $R_p$, where the GKex model
follows the behavior of the data up to the highest available values
of $Q^2$, whereas in the high-$Q^2$ regime the other models differ
substantially from the data.

Note that the BHM model is further constrained to fit
time-like data. The previous version of the GKex
model~\citep{Lomon:2002jx} was shown to provide a qualitative fit to
the time-like data by Tomasi-Gustafsson et
al.~\citep{TomasiGustafsson:2005kc} and a combined fit of the model to
space- and time-like data is underway~\citep{Pacetti:2010aa}.


\section{Insights in Momentum Space within the VMD + DR Model}
\label{subsec:insights}

\begin{figure}[htbp]
 \includegraphics[width=\textwidth]{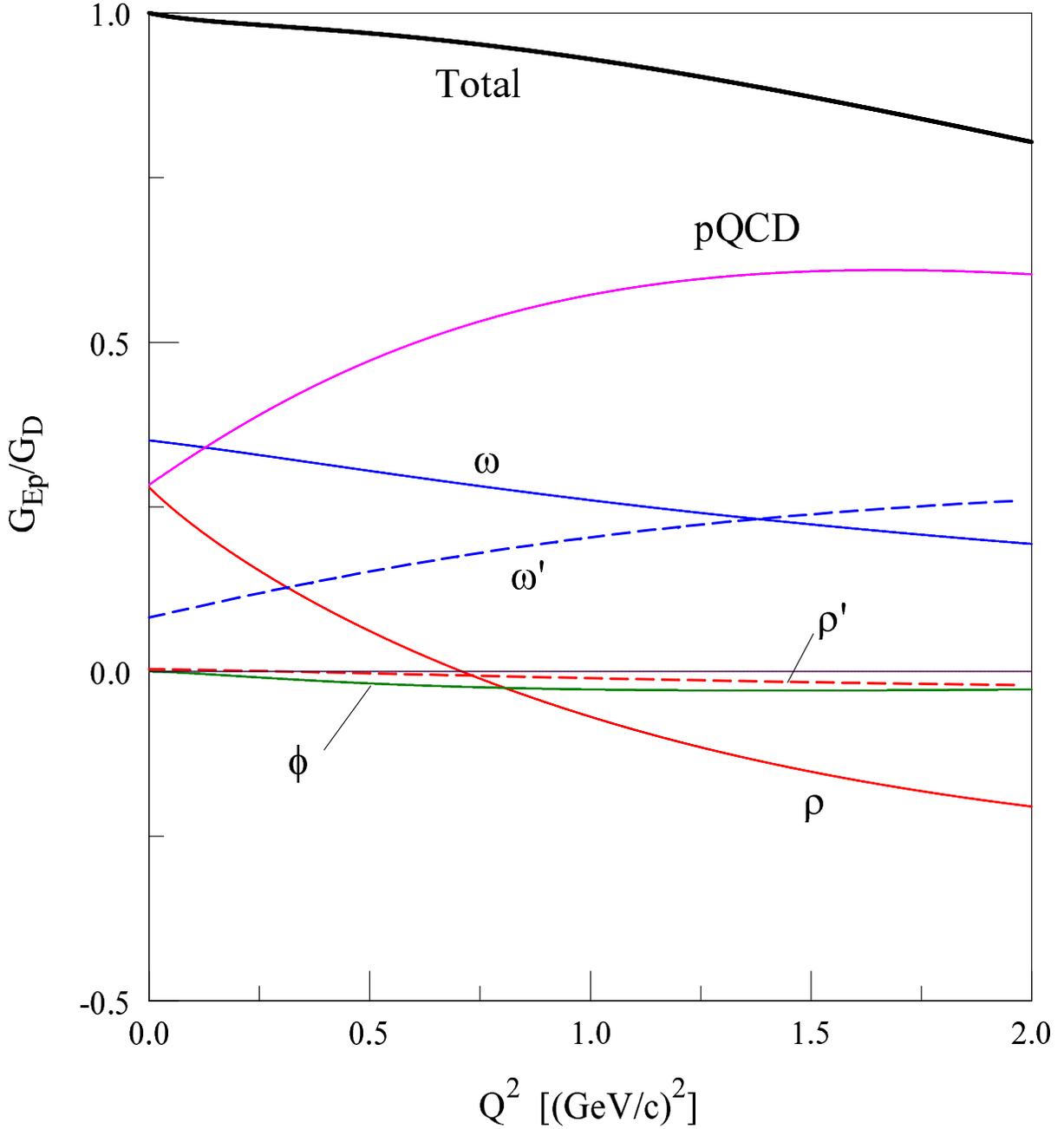}
 \caption{$G_E^p$ normalized to $G_D$ showing the relative
 contributions of the various vector mesons from the GKex model
 together with the pQCD contribution. In color online.}
 \label{twdFig1}
\end{figure}

\begin{figure}[htbp]
 \includegraphics[width=\textwidth]{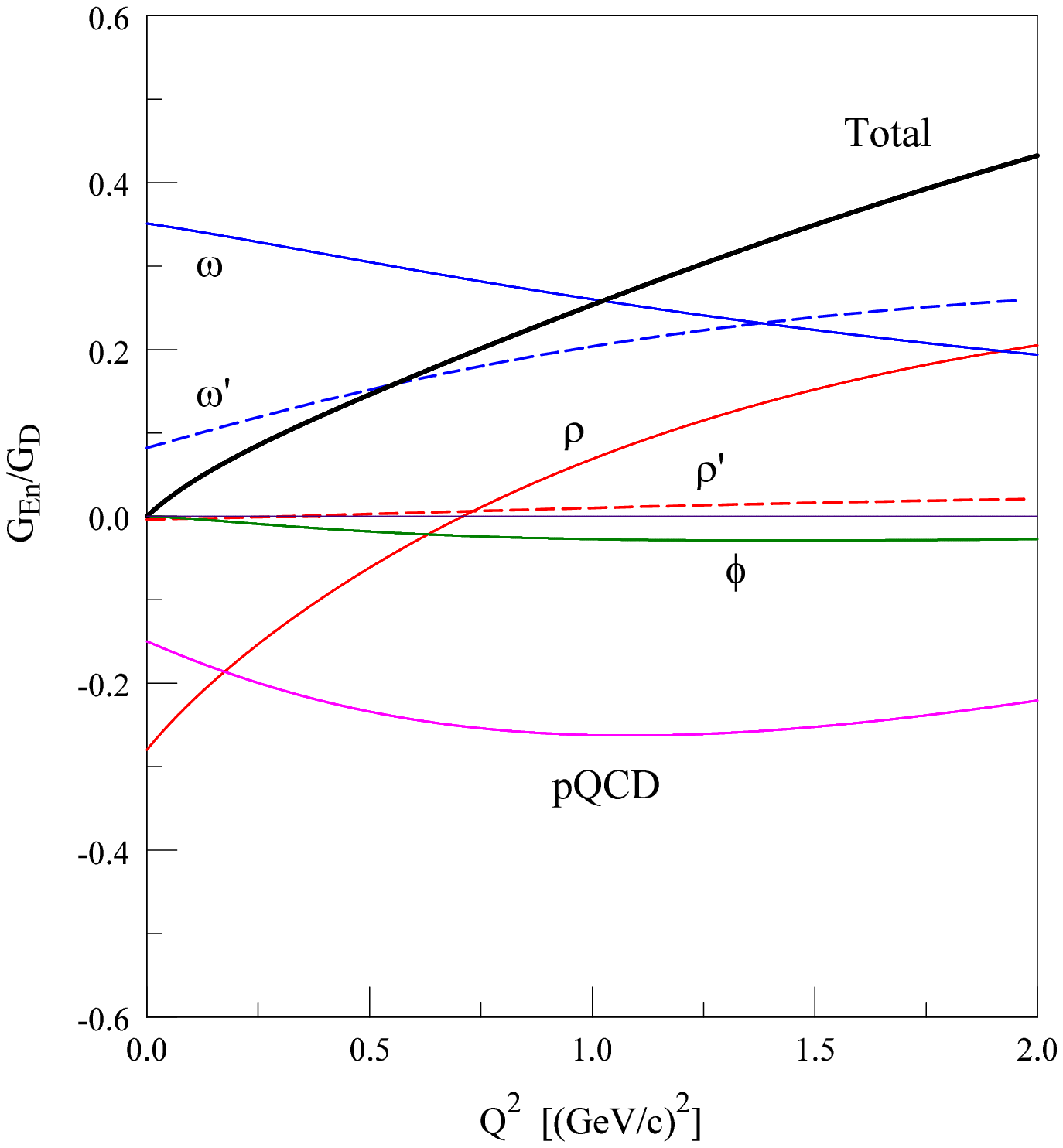}
 \caption{$G_E^n$ normalized to $G_D$ showing the relative
 contributions of the various vector mesons from the GKex model
 together with the pQCD contribution. In color online.}
 \label{twdFig2}
\end{figure}

\begin{figure}[htbp]
 \includegraphics[width=\textwidth]{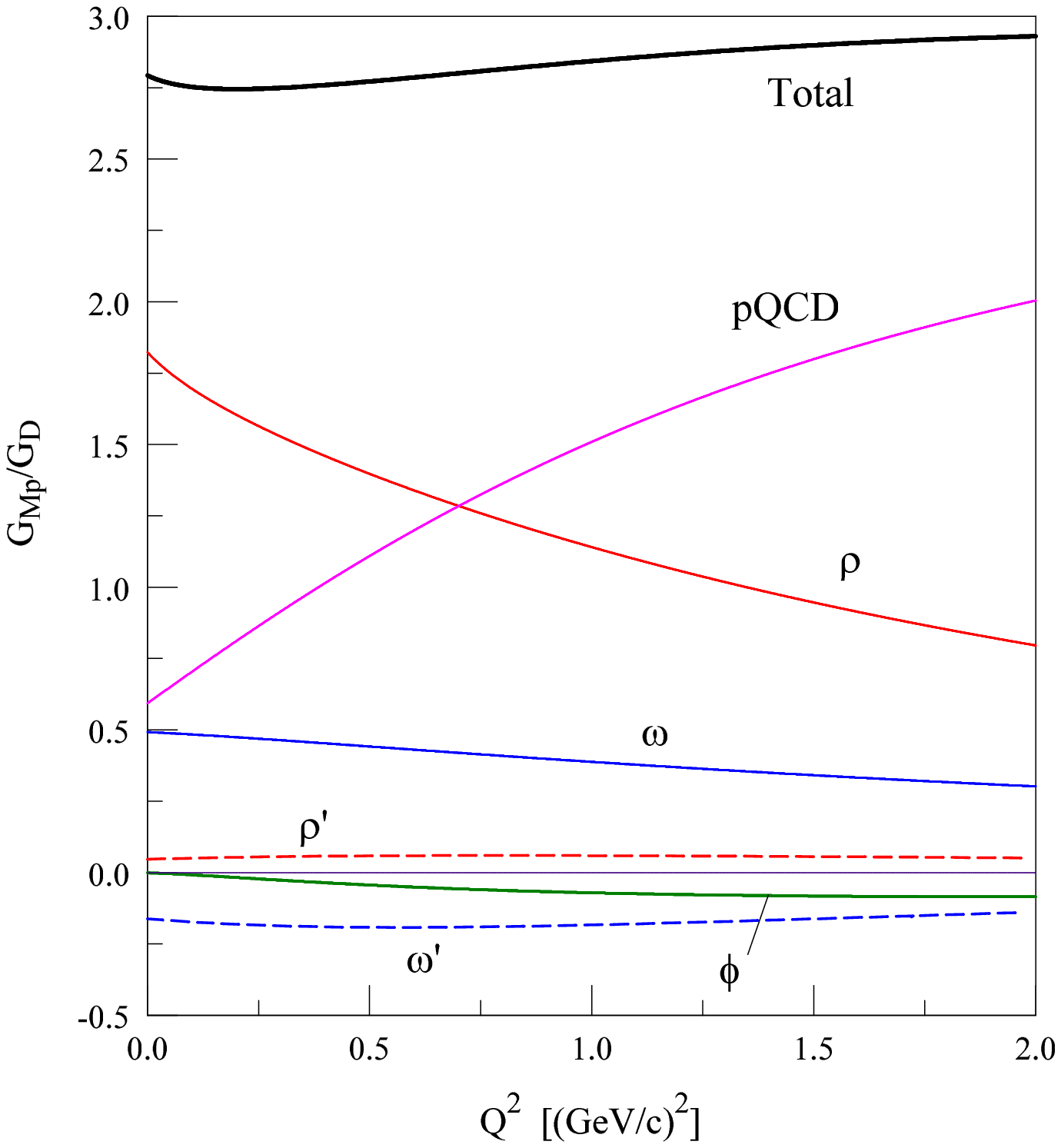}
 \caption{$G_M^p$ normalized to $G_D$ showing the relative
 contributions of the various vector mesons from the GKex model
 together with the pQCD contribution. In color online.}
 \label{twdFig3}
\end{figure}

\begin{figure}[htbp]
 \includegraphics[width=\textwidth]{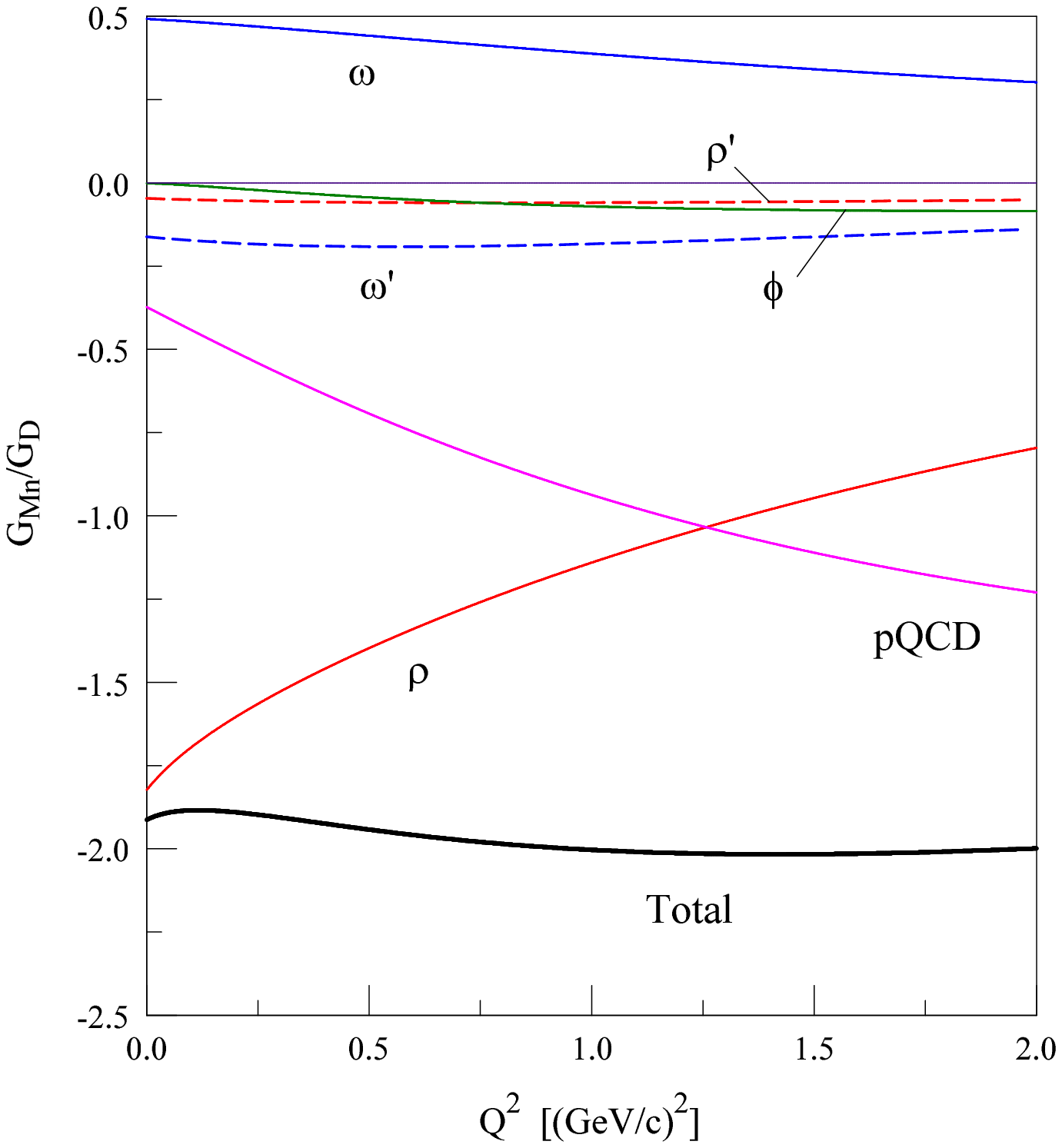}
 \caption{$G_M^n$ normalized to $G_D$ showing the relative
 contributions of the various vector mesons from the GKex model
 together with the pQCD contribution. In color online.}
 \label{twdFig4}
\end{figure}

In Figs.~\ref{twdFig1}--\ref{twdFig4} the four types of form factors
divided by the standard dipole form factor $G_D$ are shown as
functions of $Q^2$ over the range 0--2 (GeV/c)$^2$. Each is broken
down into the individual contributions from the vector mesons and
from the term that carries the asymptotic behavior, labelled pQCD.
Several insights emerge from this GKex model representation. First,
the $\phi$ and $\rho^\prime$ mesons do not play very important roles
in this region of momentum transfer for any of the four types of
form factors. Second, the $\omega^\prime$ contribution is important
for the electric form factors (Figs.~\ref{twdFig1}--\ref{twdFig2}),
but less so for the magnetic form factors
(Figs.~\ref{twdFig3}--\ref{twdFig4}). The $\rho$, $\omega$ and pQCD
contributions are important in all cases. Note that for the electric
form factors the $\rho$ has a crossing at $Q^2\sim 0.7$ (GeV/c)$^2$
which leads to interesting interplay with the other mesons, being
constructive or destructive interferences depending on the region of
momentum transfer of interest. The magnetic form factors in
Figs.~\ref{twdFig3}--\ref{twdFig4} yield a final result which is
roughly dipole in shape over the region of momentum transfer shown
in the figures (the results presented there are divided by the
dipole form factor and so being dipole corresponds to having a flat
curve). However, upon looking in more detail at the breakdown into
the individual contributions, one sees that this arises
essentially from the opposing behavior of the $\rho$ and pQCD
pieces. The $\rho$ alone, for example, is more monopole in
character, as discussed in Sect.~\ref{sec:VMD}. The compensation is
not complete, however, and the $\omega$ also plays a role in
yielding the total. This leads to the total curves being flat at
roughly the 5--8\% level. In contrast, for $G_E^p$
(Fig.~\ref{twdFig1}) the $\rho$ contribution wins and the net result
falls faster than dipole, an explicit demonstration of what all
VMD-type approaches have always predicted and now appears in the
results obtained using polarization observables, as discussed above.
Finally, for $G_E^n$ shown in Fig.~\ref{twdFig2} the situation is
even different: the $\omega$ and $\omega^\prime$ compensate almost
exactly to yield a dipole behavior, as they do for $G_E^p$, since
these are isoscalar contributions and hence the same in the two
cases; the pQCD contribution is flatter than in the other cases; and
accordingly the $\rho$ drives the rising behavior of $G_E^n/G_D$.

\begin{figure}[htbp]
 \includegraphics[width=\textwidth]{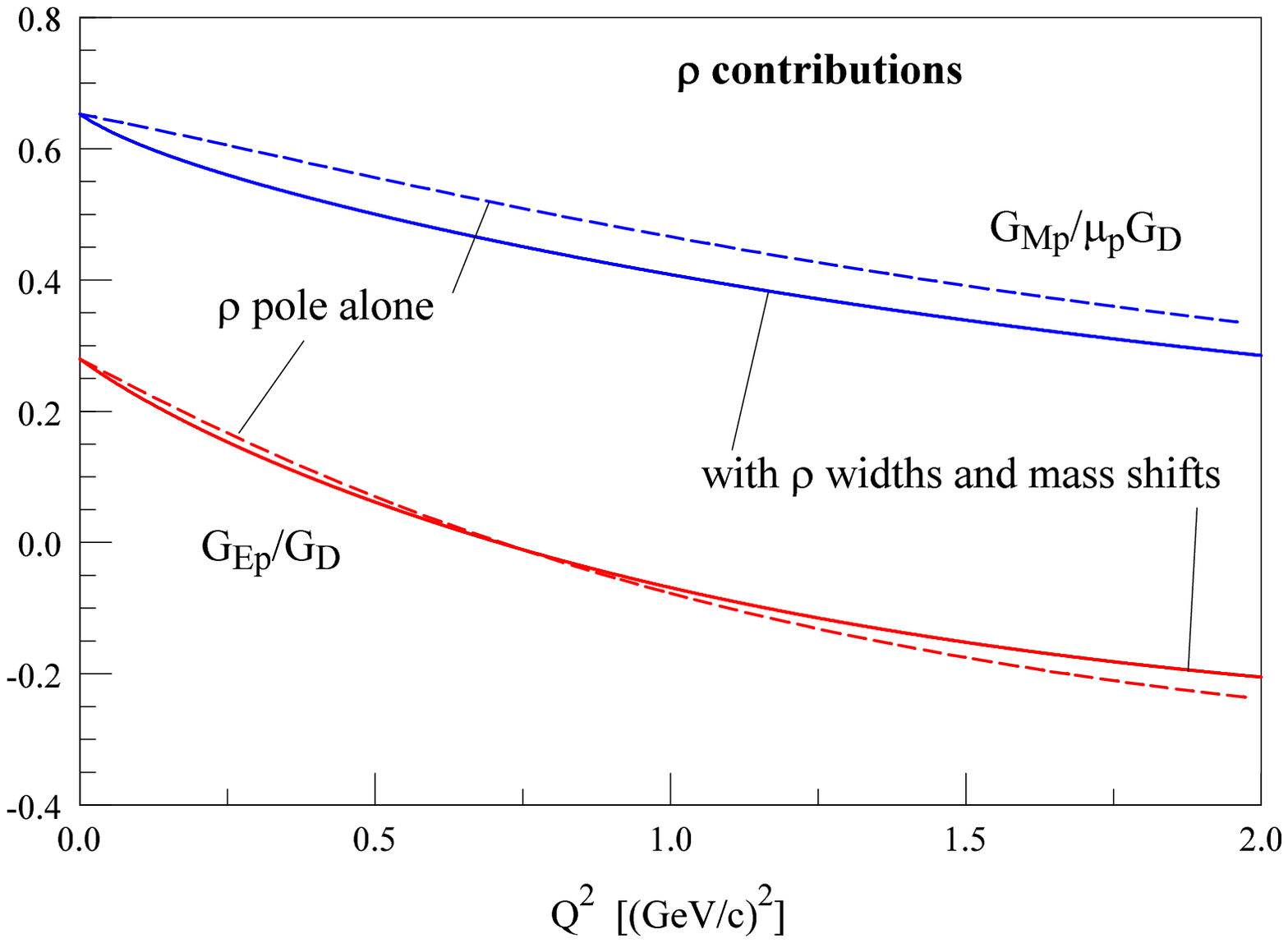}
 \caption{$G_E^p$ normalized to $G_D$ and $G_M^p$ normalized to $\mu_p
 G_D$ showing the $\rho$ contributions from the GKex model with and
 without the widths and mass shifts. In color online.}
 \label{twdFig4width}
\end{figure}

Finally, let us discuss the role of the $\rho$ width. In
Fig.~\ref{twdFig4width} the $\rho$ contributions are shown for
$G_E^p$ and $G_M^p$ (for $G_E^n$ and $G_M^n$ the results are the
same magnitude, but opposite signs, since the $\rho$ is isovector).
The solid curves repeat the results shown in Figs.~\ref{twdFig1}
and~\ref{twdFig3}, while the dashed curves display what happens when
the $\rho$ width is set to zero and the mass is set to the physical
mass of the $\rho$. In Sect.~\ref{subsec:FT} we return to see what
consequences this has for the coordinate-space representations of
the charge form factors.

\section{Representations in Coordinate Space}
\label{sec:coord}

The discussion in this section is centered on transforming both what
has been measured and the results from the GKex model for the electric
form factors into coordinate space.  Several motivations exist for
doing this:

\begin{itemize}
 \item We hope to obtain some insights into how charge is distributed
 in the nucleon;
 \item We are interested in how the various ingredients of the VMD+DR
 approach are manifested differently in coordinate space than they are
 in momentum space;
 \item In particular, we wish to explore the role played by the
 coupling to the continuum and thereby to gain some insights into, for
 instance, what roles pions play in determining the nucleon's form
 factors;
 \item When characterizing the structure in coordinate space in terms
 of some set of basis functions the correlations which occur are
 different from those that enter when doing the characterization in
 momentum space and we hope to clarify this issue.
\end{itemize}
All of these are discussed in more detail below.

In context, note that a compromise is sometimes employed, that of
Fourier transforming to coordinate space only with respect to the
transverse directions (orthogonal to the boost), but leaving the third
dimension in momentum space, thereby having a mixed
representation~\cite{Miller:2007uy}. While avoiding some of the
inevitable problems discussed below, the nucleon's properties are
harder to envision in this approach.

When choosing to represent the nucleon's properties one may choose any
frame of reference, for instance, the initial-state rest frame, the
final-state rest frame, choices in between or frames boosted to the
light-cone. Inevitably, however, the initial state, the final state or
both states must be moving and therefore boosts are required when
attempting to relate to properties in the nucleon rest frame. This
makes the problem a relativistic one.  Indeed, at high momentum
transfers this makes the interpretation in terms of coordinate-space
structure of the nucleon notoriously difficult, although at low enough
momentum transfers it may be possible to make some connections between
momentum and coordinate space. Problems occur in various guises,
depending on the approach taken; for instance, rest frame models may
be very difficult to boost and light-cone models can have troubles
when boosting from the infinite momentum frame back to physical frames
of reference.

Clearly it is important to choose the least relativistic frame of
reference to optimize one's chances. This choice is the so-called
Breit frame, as may be seen simply by minimizing the product of the
boost factors
\begin{eqnarray}
 \gamma _i  &=& E_i /m_N  \nonumber \\
 \gamma _f  &=& E_f /m_N
 \label{eq:boosts}
\end{eqnarray}
for the boosts involved in relating the moving initial and final
nucleon states to their rest frames. One has
\begin{eqnarray}
 p_f  &=&  - p_i  = q/2 \label{eq:Breitp}\\
 \omega  &=& 0\; \leftrightarrow \;\sqrt {|Q^2 |}  = |{\bf q}| \label{eq:Breitmom}\\
 \gamma _f  &=& \gamma _i  \equiv \gamma _{Breit}  = \sqrt {1 + \tau }\, ,
 \label{eq:Breitgamma}
\end{eqnarray}
that is, the resulting Breit frame has the initial- and final-state
nucleons moving with $\mp${\bf q}$/2$, where {\bf q} is the 3-momentum
of the virtual photon involved in the electron scattering process. The
energy transfer that results is zero and hence $Q^2=|${\bf q}$|^2 =
q^2$. One may then define the Breit-frame electric distributions as
the Fourier transforms
\begin{equation}
 4\pi r^2 \rho _{Breit}^{p,n} (r) \equiv \frac{2}{\pi
 }\int\limits_0^\infty  {dq\;qr\sin qr\;\left. {G_E^{p,n} (Q^2 )}
 \right|} _{Breit}\, . \label{eq:breitE}
\end{equation}
Note that this is only a definition. For the reasons mentioned above,
the resulting functions are not generally to be interpreted as the
proton and neutron charge distributions, although they are perfectly
well-defined quantities.

To obtain some feeling for where the interpretations as charge
distributions clearly should be invalid (and therefore for where they
may be reasonable) it helps to compare the Compton wavelength
$\lambda_C = \hbar c / Mc^2 \cong 0.21$~fm, where $M$ is the mass of
the nucleon, with the characteristic scale probed at a given momentum
transfer $\lambda (q) \sim \hbar c / q$. These become equal when
$q\sim 1$ GeV/c, and thus one must expect functional dependence at
even higher momentum transfers or, corresponding, smaller distance
scales to lie beyond simplistic non-relativistic intuition. At lower
momentum transfers --- corresponding to distance scales significantly
larger than the nucleon's Compton wavelength --- there may be some
validity to the interpretation of the coordinate-space distributions
as charge or spin distributions. An insightful discussion of what toy
models have to offer in this long-wavelength regime is contained
in~\citep{Isgur:1998er}.


\subsection{Insights obtained using the Breit-Frame Fourier Transform of the GKex Model}
\label{subsec:FT}

\begin{figure}[htbp]
 \includegraphics[width=\textwidth]{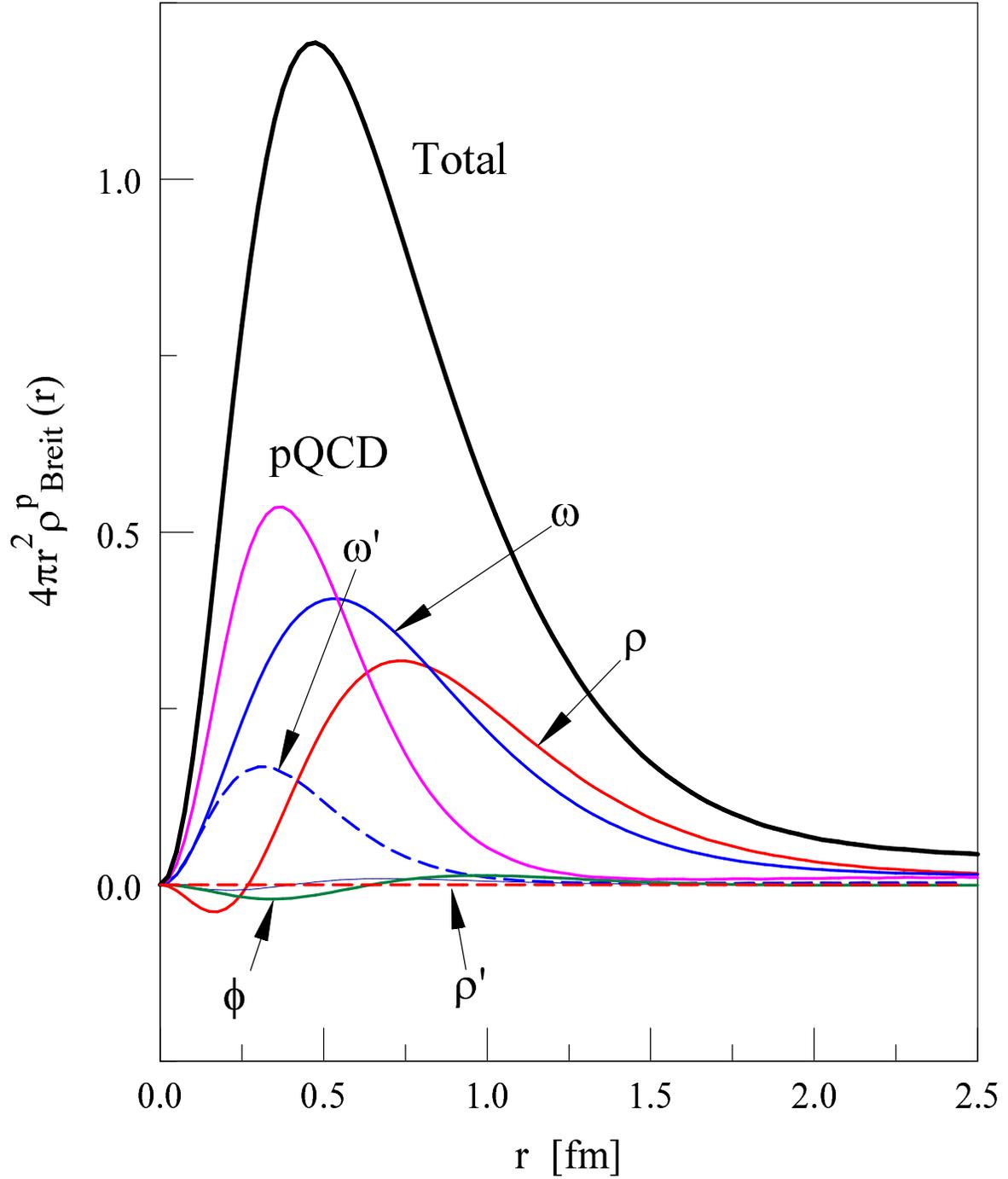}
 \caption{$4\pi r^2 \rho_{Breit}^p (r)$ showing the relative
 contributions of the various vector mesons from the GKex model
 together with the pQCD contribution. In color online.}
 \label{twdFig5}
\end{figure}

\begin{figure}[htbp]
 \includegraphics[width=\textwidth]{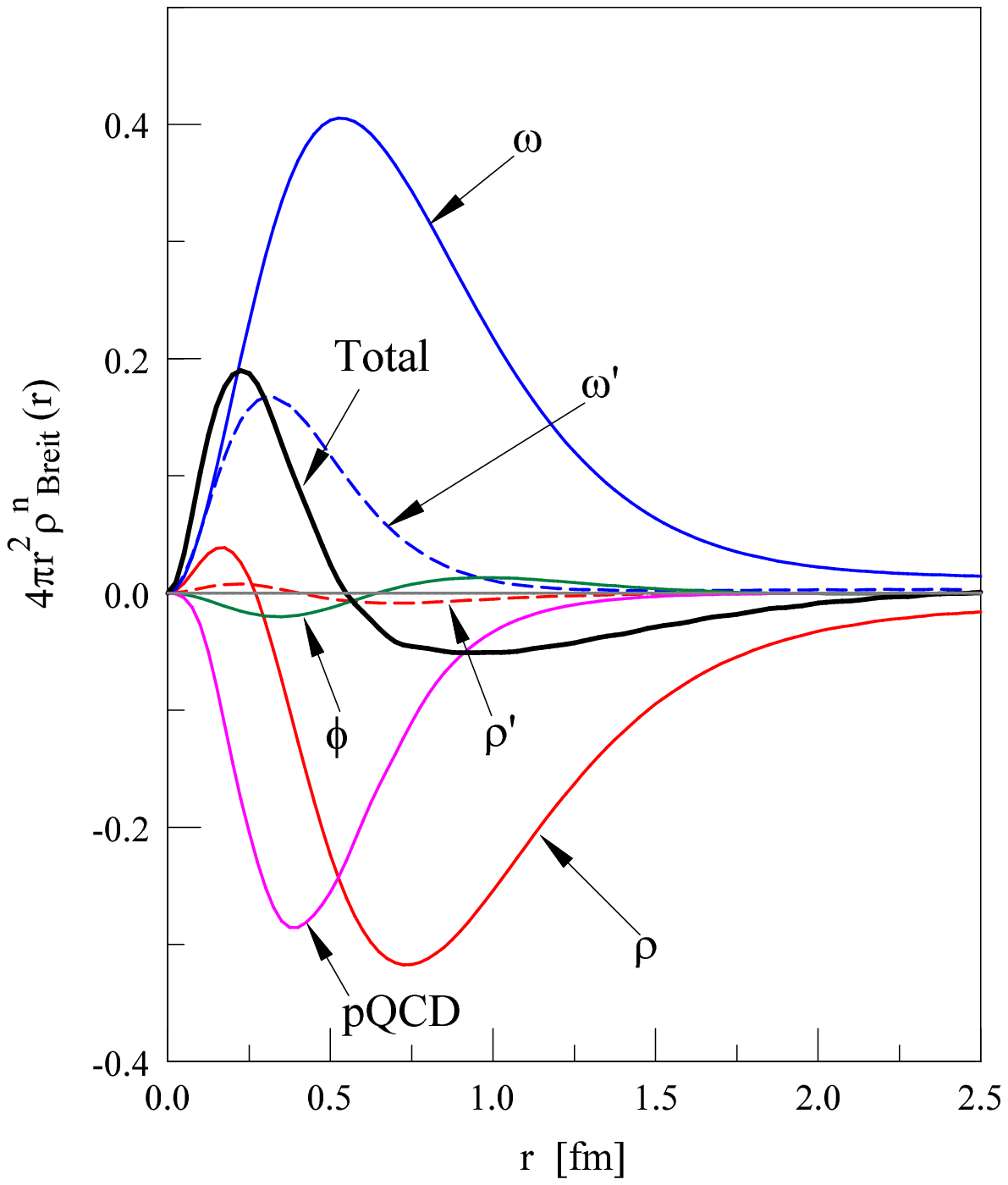}
 \caption{$4\pi r^2 \rho_{Breit}^n (r)$ showing the relative
 contributions of the various vector mesons from the GKex model
 together with the pQCD contribution. In color online.}
 \label{twdFig6}
\end{figure}

In Figs.~\ref{twdFig5}--\ref{twdFig6} we show the Breit-frame
Fourier transforms of the charge (electric) form factors of the
proton and neutron, respectively, together with the individual
contributions from the vector mesons and the asymptotic (pQCD). That
is, the figures show the Fourier transform of the GKex model results
discussed in Sect.~\ref{subsec:insights}. For the totals (the entire
GKex model form factors) one has results which integrate to 1 (0)
for the proton (neutron), since what is plotted is $4\pi r^2$ times
the Breit-frame Fourier transforms. For the neutron one sees a
positive contribution at small distances and a negative one at large
distances, which is consistent with the fact that the mean-square
radius for the neutron is $\left\langle
r^2\right\rangle_{En}=-0.115\pm0.0035$
fm$^2$~\citep{Kopecky:1995zz}. This is also consistent with a simple
picture where isovector mesons such as the $\pi$ and $\rho$ extend
to large distances and form the ``meson cloud". For example,
although unrealistically simple, a model where a neutron spends part
of its time as a ``proton + negative pion" would yield just such a
charge polarization, and not the reverse with a negative ``core" and
a positive ``cloud". Again, one is cautioned not to interpret these
distributions as charge or spin distributions, except perhaps for
their large-distance behavior. The issue of interpreting the rms
charge radius of the neutron is discussed in~\citep{Isgur:1998er}.

Let us now discuss the individual contributions in somewhat more
detail. As before the $\rho^\prime$ and $\phi$ contributions are seen
to be very small, while the rest of the contributions play important
roles. For the Breit-frame Fourier transform of $G_E^p$
(Fig.~\ref{twdFig5}) these mostly add together to form the total,
whereas for the Breit-frame Fourier transform of $G_E^n$
(Fig.~\ref{twdFig6}) the isoscalar mesons ``fight" against the
isovector mesons and the pQCD term to yield a relatively small net
result. In both cases the longest-range effects arise from the $\rho$
and next from the $\omega$, while the $\omega^\prime$ and pQCD
contributions lie at small distances. Indeed, beyond about 0.7~fm most
of the Breit-frame Fourier transform of $G_E^p$ is contained in the
$\rho$ and $\omega$ alone (the neutron case is more complicated, due
to the delicate cancellations seen in the figure).

\begin{figure}[htbp]
 \includegraphics[width=\textwidth]{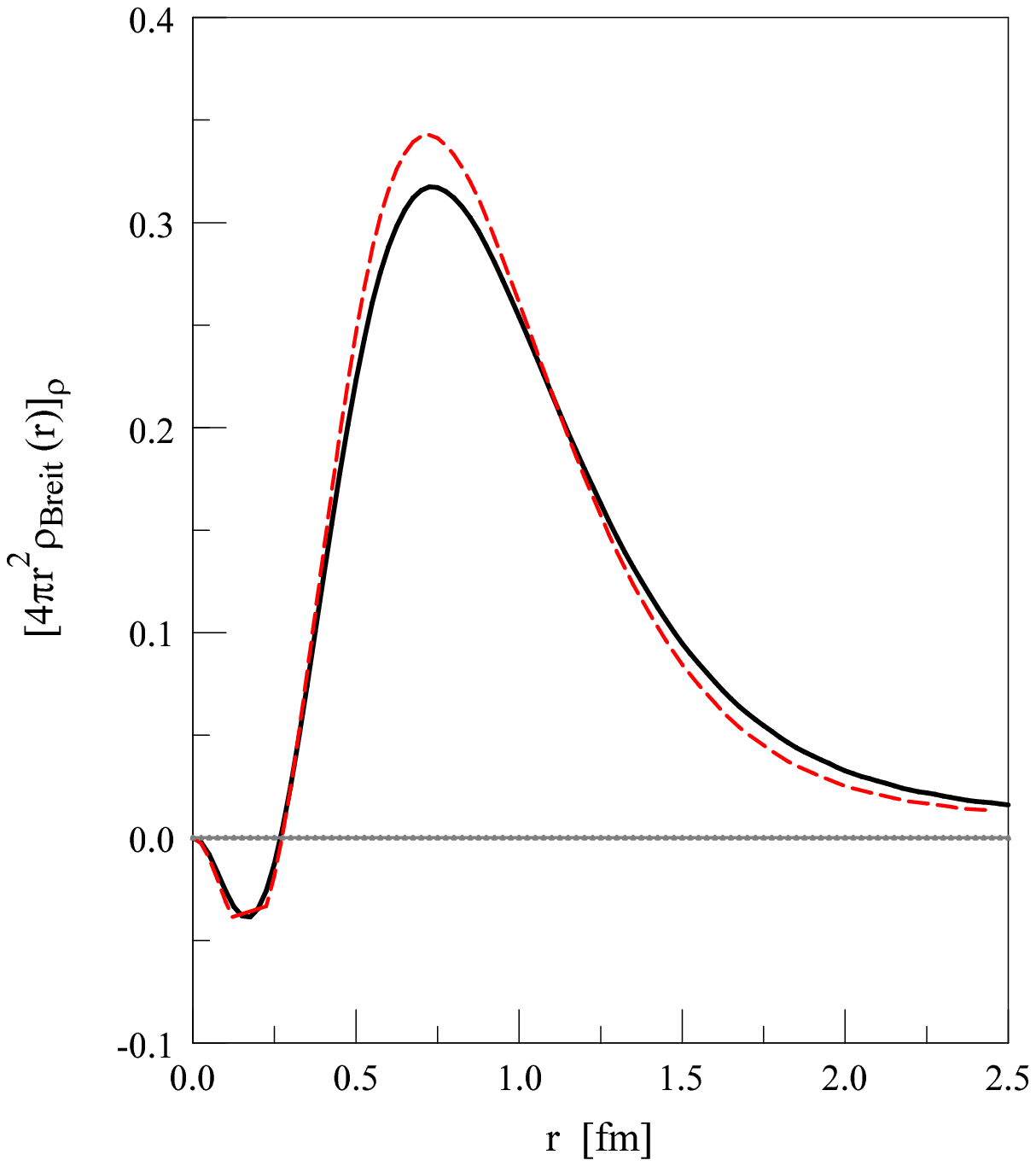}
 \caption{$4\pi r^2 \rho_{Breit}^p (r)$ showing the $\rho$-meson
 contribution from the GKex model with the width included (solid
 curve), and with the width turned off and the mass set to the
 physical $\rho$ mass (dashed curve). In color online.}
 \label{twdFig7}
\end{figure}

The effect of ``turning off" the $\rho$ width was discussed in
Sect.~\ref{subsec:insights} for the momentum-space GKex model
results. Here we consider the Breit-frame Fourier transform as well.
In Fig.~\ref{twdFig7} curves are shown for the $\rho$ contributions
in the proton both with the width included (solid curve, as in
Fig.~\ref{twdFig5}) and with it set to zero and the mass of the
$\rho$ set to its physical value (dashed curve). The latter is seen
to have a bit more strength at smaller distances, although the
effect is not pronounced. In the GKex representation of the form
factors the only place that contributions from pions appear
explicitly is via the width the $\rho$ takes on, {\it i.e.} through
connections to the $\pi\pi$-continuum. Otherwise only vector mesons
and the asymptotic form occur in the model. Thus, turning off these
$\rho$-width contributions effectively eliminates explicit pions
from the problem, and one must conclude that the latter are
relatively unimportant.

\subsection{Results in Coordinate Space}
\label{subsec:coordresults}

\begin{figure}
  \centering
  \includegraphics[width=\textwidth]{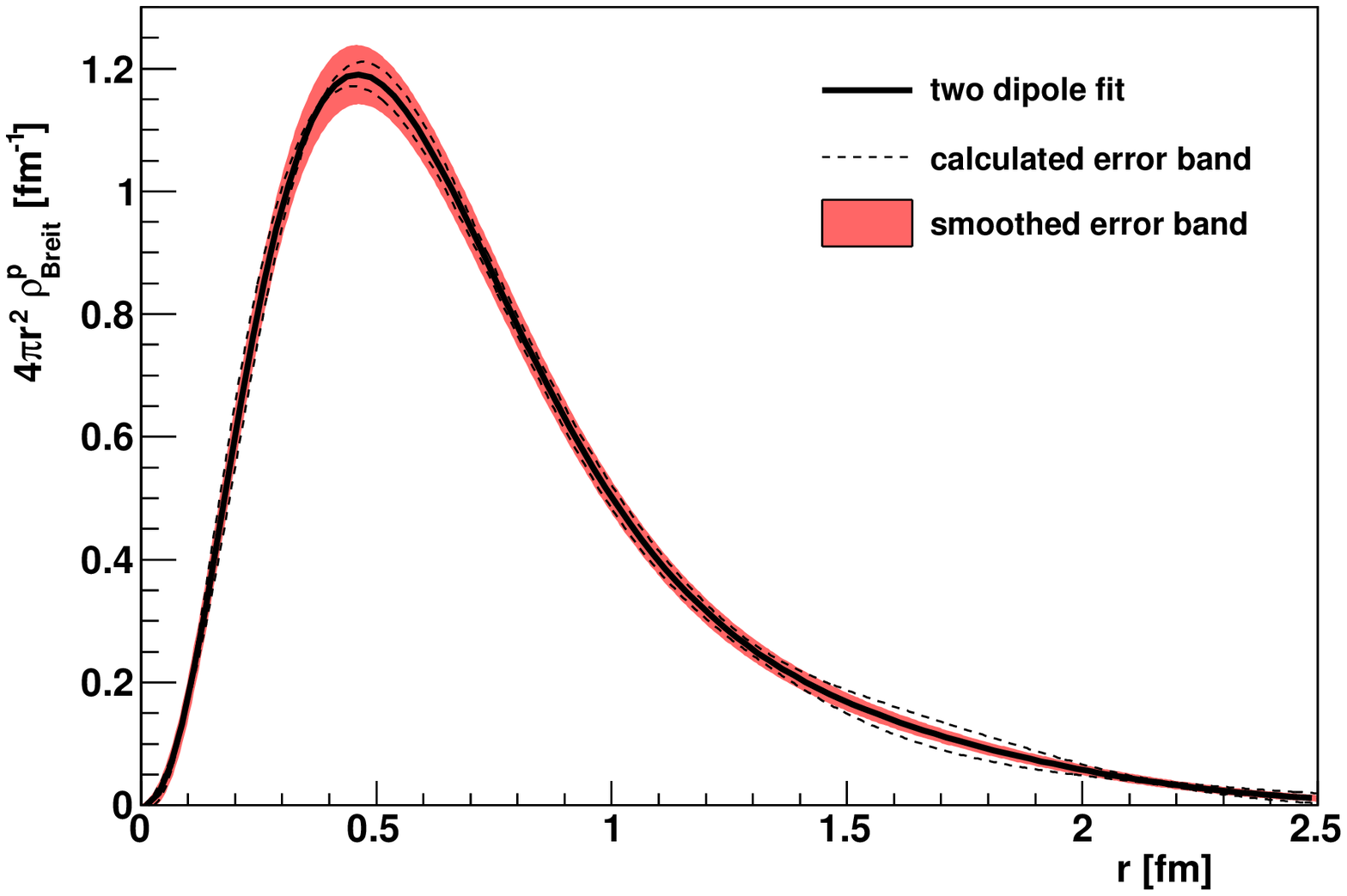}
  \caption{Breit frame Fourier transform of $G_E^p$, as
  appeared in the long range plan~\cite{:2008jna}, with both the
  calculated and smoothed error bands. In color online.}
  \label{fig:lrpp}
\end{figure}

\begin{figure}
  \centering
  \includegraphics[width=\textwidth]{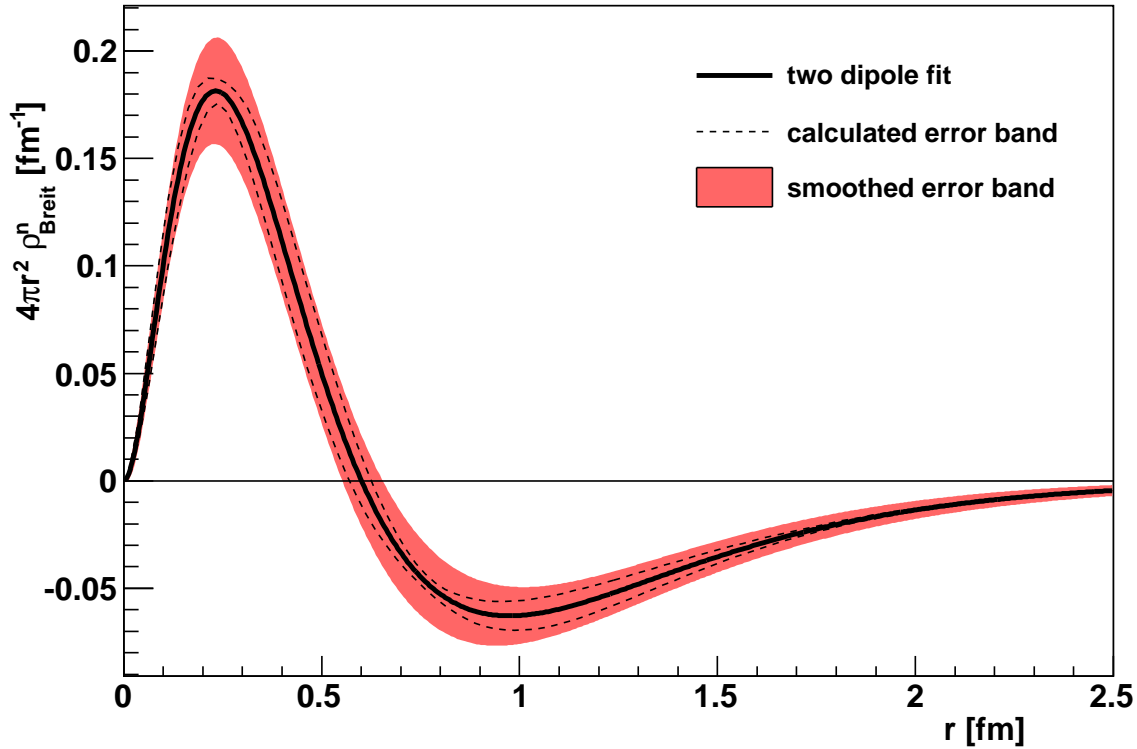}
  \caption{Breit frame Fourier transforms of $G_E^n$, as
  appeared in the long range plan~\cite{:2008jna}, with both the
  calculated and smoothed error bands. In color online.}
  \label{fig:lrpn}
\end{figure}

Again, given the caveats discussed in the introduction to this
section, the world data for $G_E^{p,n}$ may be Fourier-transformed
using Eq.~(\ref{eq:breitE}).  In order to obtain Fourier transforms of
the experimental data, the world data of $G_E^p$ and $G_E^n$ were fit
to various parameterizations which were then transformed numerically.
Earlier work presented in the DOE/NSF NSAC Long Range
Plan~\cite{:2008jna} was based on the data and parameterization used
in~\cite{Crawford:2006rz,Geis:2008ha}.  For the proton, this was the
6-parameter phenomenological fit function of
\cite{Friedrich:2003iz} fit to the data from~\cite{Simon:1980hu,
Ron:2007vr, Price:1971zk, Berger:1971kr, Hanson:1973vf,
Pospischil:2001pp, Milbrath:1997de, Dieterich:2000mu, Jones:1999rz,
Gayou:2001qd, Crawford:2006rz}.  For the polarized data, $G_E^p$ was
obtained by combining the form factor ratio with the
Kelly~\cite{Kelly:2004hm} fit of $G_M^p$.  For the neutron the fit
function was reduced to the sum of two dipoles, fit to the data
of~\cite{Eden:1994ji, Herberg:1999ud, Ostrick:1999xa, Madey:2003xk,
Seimetz:2005vg, Glazier:2004ny, Passchier:1999cj, Zhu:2001md,
Warren:2003ma, Becker:1999tw, Bermuth:2003qh, Geis:2008ha}.  The
charge of the neutron was constrained to zero, leaving three free
parameters.  The RMS charge radius squared of~\cite{Kopecky:1995zz}
was included in the fit as an extra datum, not as a constraint.
Figs.~\ref{fig:lrpp} and~\ref{fig:lrpn}
show the Fourier transforms of these fits.

The error bands in Figs.~\ref{fig:lrpp} and~\ref{fig:lrpn} were
obtained by combining the variation from each fit parameter with
the full covariance matrix.  The calculated error bands, shown
with dotted lines, have large oscillations in width, even dropping
to $\delta\rho_{Breit} \sim 0$ around $r$=0.37~fm for the proton
and $r$=0.75~fm for the neutron.  The calculated uncertainty for
the proton also gets significantly smaller around $r$=0.75~fm.
This is clearly model dependence: the Fourier transform of this
particular model has no flexibility a that point to respond to
variations in the data.  The shaded error bands in
Figs.~\ref{fig:lrpp} and~\ref{fig:lrpn}, were smoothed out to
account for the model dependence, producing the error bands, which
appeared in \cite{:2008jna}.

This surprising behavior illustrates an interesting point, that a
family of curves which fit the data well in momentum space may contain
very little information or coverage of coordinate space.  In choosing
an appropriate model, one typically searches for the smoothest family
of curves that fit the data with a reasonable $\chi^2$.  In contrast,
the Fourier transform inherently includes information on all
frequencies, not just smooth low frequencies.  For example, the fit to
a constant function $f(k)=a$ only determines a single point at the
origin of the Fourier transform $\tilde f(x)= a\delta(x)$.  Even
arbitrary fit functions in one parameter can often be approximated by
$f(k)=g(k)+a$ for a fixed function $g(k)$.  In momentum space, that
function will have a uniform error-band over the entire domain, but
that error is completely correlated along the entire function.  The
Fourier transform has non-zero error bars only at the origin in
position space.

To obtain a reasonable Fourier transform with meaningful error bands,
it is necessary to fit a function which spans both position and
momentum space.  This can be done by expanding the form factors in an
orthogonal set of basis functions $\sum_{n=0}^N \tilde f_n(k)$, using
the simple prescription $Q^2 =
\hbar^2 k^2$.  The kernel of the Fourier transform is unitary, ensuring an
expansion $\sum_{n=0}^N f_n(r)$ in orthogonal basis functions in
position space also.  Following Kelly~\cite[Eqs.~(28,
29)]{Kelly:2002if}, we fit the data to two orthogonal basis functions.
The first is the Fourier-Bessel Expansion (FBE), the wave functions of
an infinite spherical well of radius $R_{max}$ in position space,
\begin{eqnarray}
  \label{eq:FBE}
  f_n(r) &=& j_0(k_n r) \varTheta(R_{max}-r), \\
  \tilde f_n(k) &=& \frac{(-1)^n R_{max}}{k^2-k_n^2}j_0(k R_{max}). \\
\end{eqnarray}
These functions are localized in frequency, peaking at $k_n =
n\pi/R_{max}$, with a hard cutoff at the $n$-th zero of the $l=0$
spherical Bessel function $j_0(x)$ at $R_{max}$.  The second is the
Laguerre-Gaussian expansion (LGE), the wave functions of a spherical
harmonic oscillator of frequency $\omega = 2\hbar/m b^2$ for fixed
parameter $b$,
\begin{eqnarray}
  \label{eq:LGE}
  f_n(r) &=& e^{-x^2}L_n^{1/2}(2x^2), \\
  \tilde f_n(k) &=& \frac{\sqrt{\pi}}{4}b^3(-1)^ne^{-y^2}L_n^{1/2}(2y^2), \\
\end{eqnarray}
where $x=r/b$, $y=kb/2$, and $L_n^{1/2}$ is a generalized Laguerre
function.  These functions are localized in neither position nor
momentum.  The width of the basis functions is not fixed in coordinate
space, but increases with $n$ as $b\sqrt{n}$.  Higher-order functions
emphasize larger values of both $r$ and $k$.  These two basis sets
have quite complementary features; so it should be clear by comparing
results from the two expansions which parts depend on the particular
basis set used and which are model-independent.  In this paper,
relativistic corrections to the form factors or to $Q^2=\hbar^2 k^2$
are not considered as they were in~\cite{Kelly:2002if}.

There are a number of sources of uncertainty in the fits, which are
interrelated.  The maximum value of $Q^2$ of the data limits the
maximum number, $N$, of basis functions which can be fit for fixed
$R_{max}$ or $b$.  The $Q^2$ range of each basis function depends on
$R$ or $b$ so a larger number of basis functions can be used by
increasing the size of the box.  However the box size is limited by
the $Q^2$ gaps in the $G_E^p$ and $G_E^n$ database.  With the
appropriate box size, $N$ is ultimately limited by the finite number
of form factor measurements at independent values of $Q^2$.  If one
tries to use more basis functions, the fit parameters will become
highly correlated, manifest by a large error band.  Even below this
limit, as $N$ increases there are fewer data per fit parameter, and so
the error should grow as $\sqrt N$.  This increase in error is offset
by the extra information obtained in higher spatial frequencies.  The
truncation error from omitting higher frequencies is represented below
with a horizontal error bar of width $\delta r=h/4\sqrt{Q^2}$, a
quarter wavelength of the highest frequency basis function.  This is
an overestimate, since the form factors fall off rapidly with $Q^2$.

With the small number of basis functions (N=7--8) afforded by the
data, it is difficult to obtain convergence to $G^{p,n}_E(Q^2)$.
Better convergence can be obtained while retaining the model
independence by fitting only the residual form factors after
subtracting an arbitrary base function which reproduces the general
features of the data.  We used the GKex model as the base function.
The FBE or LGE expansion is used to fit the small correction to GKex
from the data, and mainly to calculate the model-independent error
band of the Fourier transform.  The quality of the base function can
be assessed by comparing the residual fit with the size of the error
band.  The model independence can be shown by comparing the FBE and
LGE expansions, and by using different base functions.

In general, the widths of the error bands of the fits to $4\pi r^2
\rho^{Breit}(r)$ were linear in $r$, superimposed with an oscillation due to
truncation after a finite number of the basis functions.  The
oscillations were approximately the frequency of the highest basis
function.  The linear part was consistent between the FBE and LGE
residual fits, but not the oscillations.  The oscillations were small
for reasonable values of $N$, but started to dominate as too many
basis functions were used.  Only the linear part of the error bands
were used in the final plots.

The complete procedure used to determine the optimal values of the
non-fit parameters $(Q^2_{max}, N, R_{max})$ or $(Q^2_{max}, N,
b)$ in the Fourier transform of the data is as follows.  The
residual $G^{p,n}_E(Q^2)$ data after subtracting the GKex model
were fit to a series of $N$ basis functions, either FBE or LGE.
The width of the error band was fit to the linear function
$\delta\rho(r) = \rho_1 \cdot r /$ 1~fm, and then $\rho_1(N)$ was
plotted as a function of the number of basis functions used each
fit.  A series of such plots $\rho_1(N; Q^2_{max})$ was generated
for data subsets with different cuts of the form
$0<Q^2<Q^2_{max}$.  The values $Q^2_{max}$ = 0.1, 0.4, 0.7, 1.0,
1.5, 2.0, 3.0, and 6.0~(GeV/c)$^2$ for the proton and $Q^2_{max}$
= 0.2, 0.3, 0.5, 1.0, and 1.5~(GeV/c)$^2$ for the neutron were
used to generate the series of plots.  At small $N$,
$\delta\rho(N)$ was the same for each value of $Q^2_{max}$.  As
$N$ increased, $\delta\rho(N)$ began to diverge for data sets with
lower values of $Q^2_{max}$.  The threshold of $N$ where the fits
began to diverge indicated the maximum number of basis functions
feasible for each $Q^2$ range, $N(Q^2_{max})$.  The $Q^2$ range
was fixed at $Q_{max}$ = 1.5~(GeV/c)$^2$ for comparison of
$\rho_{Breit}^p(r)$ and $\rho_{Breit}^n(r)$ and to avoid issues of
two photon contributions. The entire procedure was repeated with
different box sizes $R_{max}$ (FBE) or $b$ (LGE).  The values
$R_{max}$ and $b$ were chosen to minimize
$\delta\rho_1(N,Q^2_{max})$.  As one would expect, the optimal box
size was the same for the proton and neutron.  The best value of
$R_{max}$ was the same as in Kelly~\cite{Kelly:2002if}; however,
the best value for $b$ was about twice as large.  The parameters
obtained using this procedure are listed in
Table~\ref{tab:nonfit_params}.
\begin{table}[ht]
 \centering
 \begin{tabular}{c@{\qquad}c@{\qquad}c@{\qquad}c@{\qquad}c}
   \hline
   FF     & $Q^2_{max}$       & $N$    & $R_{max}$ & $b$ \\
   \hline
   $G_E^p$  & 1.5~(GeV/c)$^2$ & 8    & 4~fm & 1.05~fm \\
   $G_E^n$  & 1.5~(GeV/c)$^2$ & 7    & 4~fm & 1.11~fm \\
   \hline
 \end{tabular}
 \caption{Fixed parameters used in the fits of FBE and LGE basis function to
   the residual $G^{p,n}_E(Q^2)$ after subtracting the GKex model.}
 \label{tab:nonfit_params}
\end{table}

\begin{figure}[htbp]
 \includegraphics[width=\textwidth]{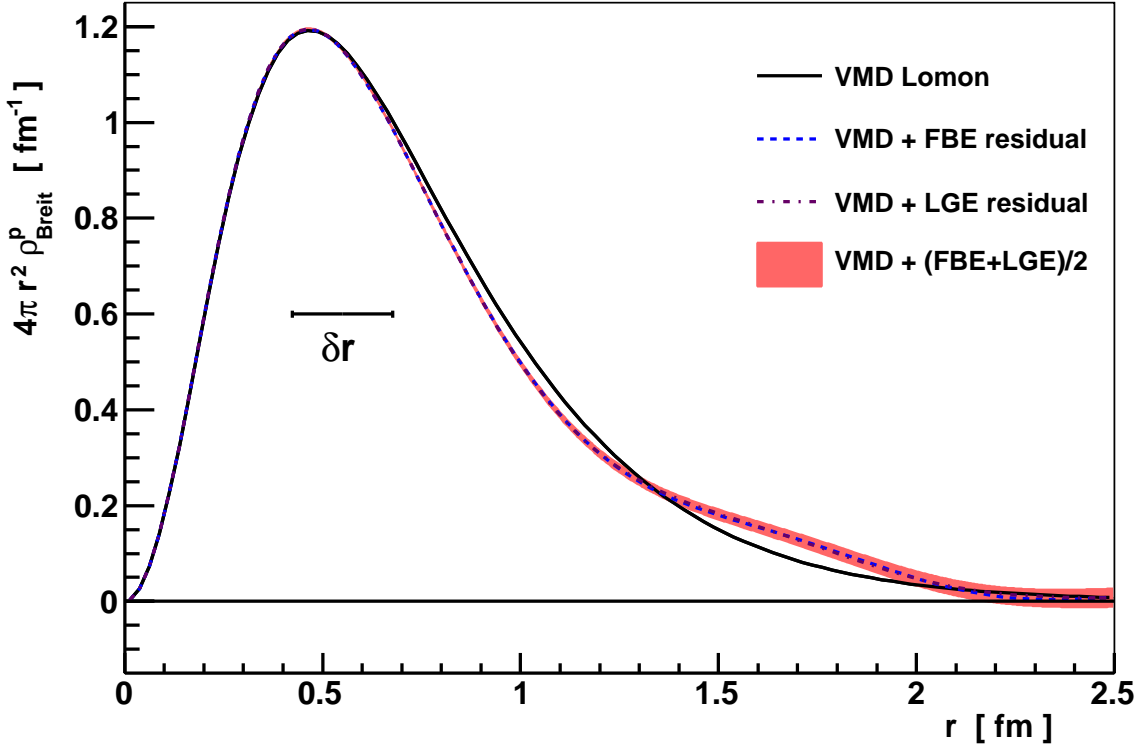}
 \caption{Coordinate-space representation
 $4\pi r^2\rho^p_{\mbox{\tiny Breit}}(r)$
 obtained using Eq.~(\ref{eq:breitE}) with $G_E^p(Q^2)$ together with
the GKex VMD model of Lomon~\cite{Lomon:2006sf}. In color online.}
 \label{fig:rhoEp}
\end{figure}

\begin{figure}[htbp]
 \includegraphics[width=\textwidth]{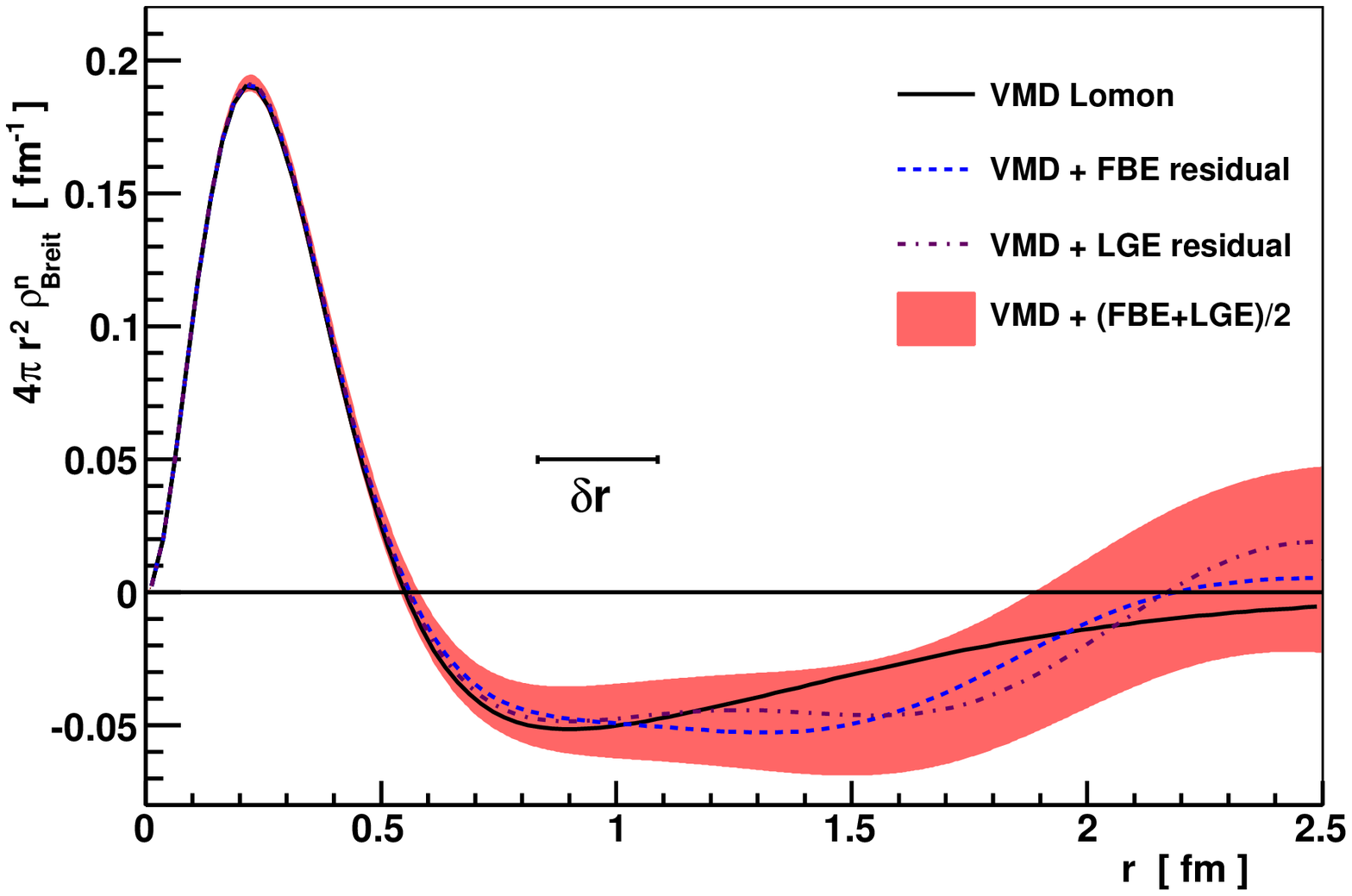}
 \caption{Coordinate-space representation
 $4\pi r^2\rho^n_{\mbox{\tiny Breit}}(r)$
 obtained using Eq.~(\ref{eq:breitE}) with $G_E^n(Q^2)$. In color online.}
 \label{fig:rhoEn}
\end{figure}

In Figs.~\ref{fig:rhoEp} and \ref{fig:rhoEn}, $\rho_{Breit}(r)$ for
the GKex model is compared with fits to the world data with smoothed
error bars obtained through the above procedure.  The differences
between the solid curve and the two other curves are the LGE and FBE
residuals fitted to the data. The residuals are small, but
statistically significant.  Although they deviate from the GKex
model, the FBE and LGE residuals are consistent with each other
within error. This is an important confirmation of the
model-independence of the residual fit, since the two basis
functions are very different, as described above.  To check for
coverage of the basis functions, fits to the residuals of different
parameterizations such as the F-W or two dipole forms from above
were compared with GKex+FBE and GKex+LGE and found to be consistent
within error.  We conclude that the Fourier transforms of $G_E^p$
and $G_E^n$ world data are robust with realistic error bands. To
place these Breit-frame distributions in context with other work,
note that when representing results in the light-cone frame, for
instance, that different (but not incompatible) behaviors may
emerge, showing that one's perceptions must be keyed to what frame
of reference is chosen. Examples of this type may be found in the
work of \cite{Pasquini:2006ib,Pasquini:2007iz,Miller:2008jc} where
the light-cone-frame neutron distribution may even be negative at
the origin.

\begin{figure}[htbp]
 \includegraphics[width=\textwidth]{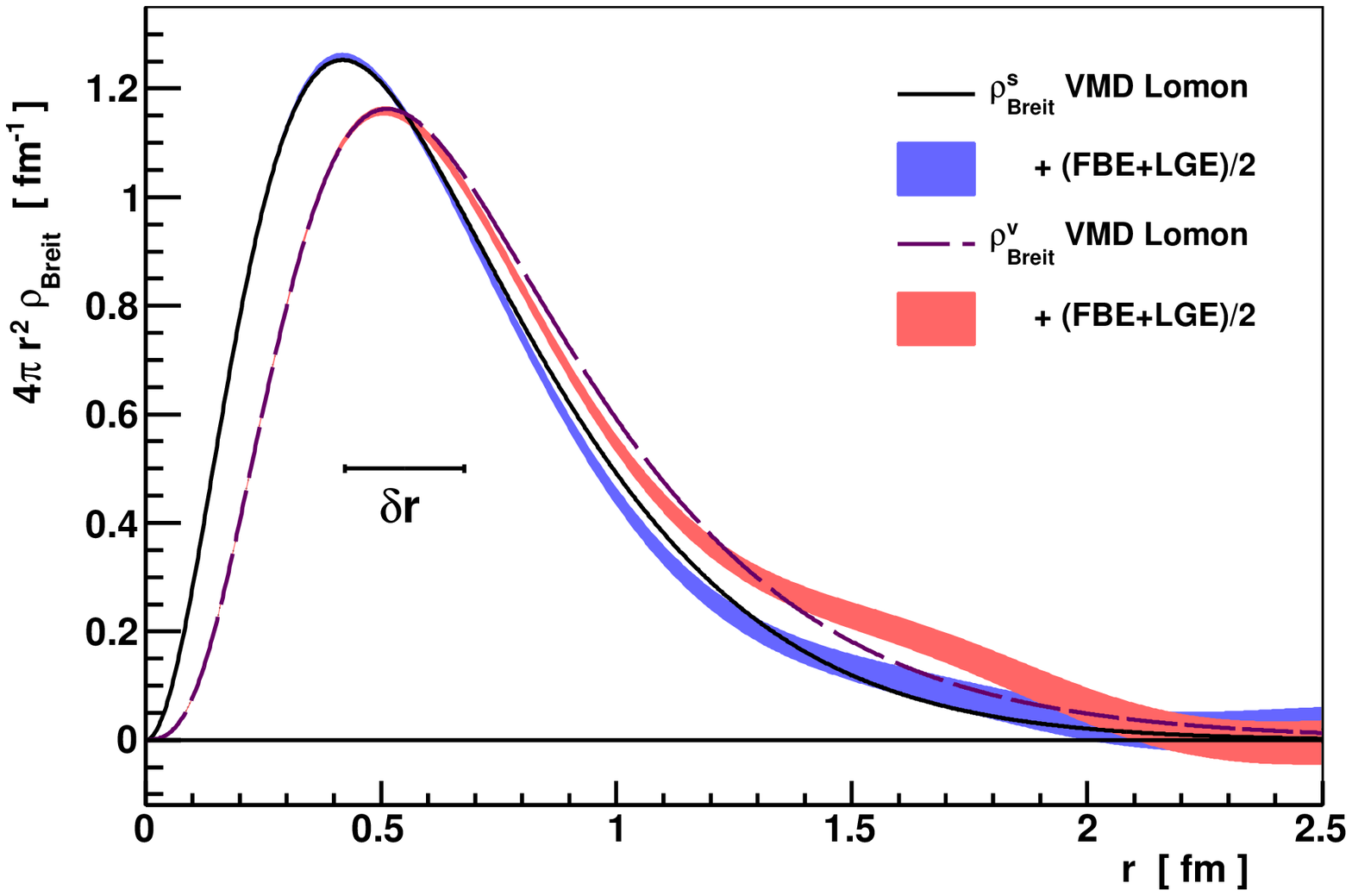}
 \caption{Isoscalar and isovector coordinate-space electric
 Breit-frame
 distributions obtained using Eqs.~(\ref{eq:isosE})
and~(\ref{eq:isovE}). In color online.}
 \label{fig:isosvE}
\end{figure}

The Breit coordinate-space electric distributions discussed above may
be combined to yield two different quantities. First, by taking sums
and differences the isoscalar and isovector Breit-frame electric
distributions shown in Fig.~\ref{fig:isosvE}
may be constructed:
\begin{eqnarray}
 \rho^s_{\mbox{\tiny Breit}}(r)  &\equiv&
 \rho^p_{\mbox{\tiny Breit}}(r) + \rho^n_{\mbox{\tiny Breit}}(r)
 \label{eq:isosE} \\
 \rho^v_{\mbox{\tiny Breit}}(r)  &\equiv&
 \rho^p_{\mbox{\tiny Breit}}(r) - \rho^n_{\mbox{\tiny Breit}}(r).
 \label{eq:isovE}
\end{eqnarray}

Since the neutron electric distribution shown in Fig.~\ref{fig:rhoEn}
is positive at small distances and negative at large distances one
sees that the isovector distribution lies outside the isoscalar one,
apparently consistent with isovector mesons playing an important role
in determining the large-distance behavior (compare
Fig.~\ref{fig:isosvE} with Figs.~\ref{twdFig5} and~\ref{twdFig6} where
one sees the $\rho$ contribution extending beyond the $\omega$
contribution).

\begin{figure}[htbp]
 \includegraphics[width=\textwidth]{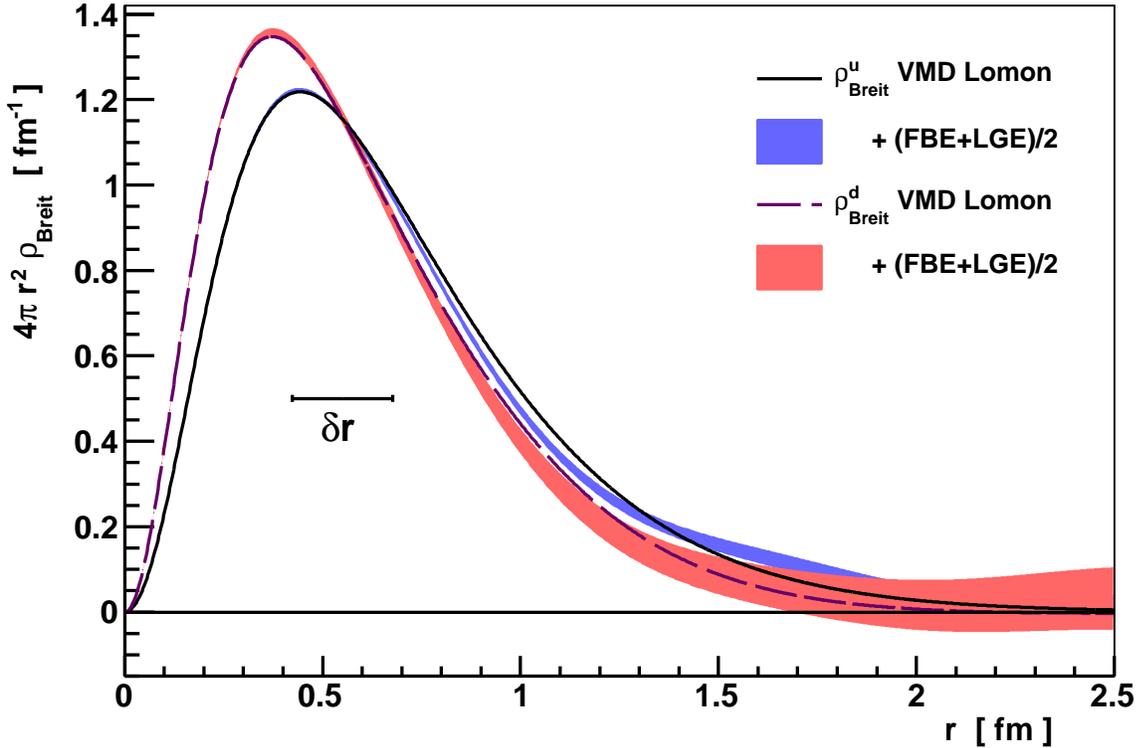}
 \caption{Up and down quark coordinate-space Breit-frame electric
 distributions obtained using Eqs.~(\ref{eq:up}) and~(\ref{eq:down}). In color online.}
 \label{fig:udE}
\end{figure}

Secondly, note that the proton and neutron Breit-frame electric
distributions may be written in terms of Breit-frame electric up and
down quark distributions (neglecting strange quark
contributions), involving the appropriate numbers of quarks (1 or
2) and quark charges (-1/3 and 2/3), both for the proton and for the
neutron:
\begin{eqnarray}
 \rho^p_{\mbox{\tiny Breit}}(r)  &\equiv& 2[\frac{2}{3}
 \rho^u_{\mbox{\tiny Breit}}(r)] + [-\frac{1}{3} \rho^d_{\mbox{\tiny
Breit}}(r)] \label{eq:proton} \\
 \rho^n_{\mbox{\tiny Breit}}(r)  &\equiv& 2[-\frac{1}{3}
 \rho^u_{\mbox{\tiny Breit}}(r)] + [\frac{2}{3} \rho^d_{\mbox{\tiny
Breit}}(r)]. \label{eq:neutron}
\end{eqnarray}
Here $\rho^u$ ($\rho^d$) denote up (down) quark distributions in
the proton; by charge symmetry these are assumed to be the same as
the down (up) quark distributions in the neutron to obtain
Eq.~(\ref{eq:neutron}), {\it i.e.} we have assumed that
\begin{eqnarray}
 \rho^u &\equiv& \rho^{u(p)} = \rho^{d(n)} \label{eq:uquark} \\
 \rho^d &\equiv& \rho^{d(p)} = \rho^{u(n)}\, . \label{eq:dquark}
\end{eqnarray}
Inverting, one may construct the corresponding up and down quark
distributions in terms of the proton and neutron distributions
\begin{eqnarray}
 \rho^u_{\mbox{\tiny Breit}}(r)  &\equiv& \rho^p_{\mbox{\tiny
Breit}}(r) + \frac{1}{2} \rho^n_{\mbox{\tiny Breit}}(r) \label{eq:up} \\
 \rho^d_{\mbox{\tiny Breit}}(r)  &\equiv& \rho^p_{\mbox{\tiny
Breit}}(r) + 2 \rho^n_{\mbox{\tiny Breit}}(r), \label{eq:down}
\end{eqnarray}
shown in Fig.~\ref{fig:udE}.


\section{Conclusions}
\label{sec:concl}

The goal of the present study has been to gain insight into the
roles played by mesons in the electromagnetic form factors of the
nucleon. A basic reference model, the GKex model of Lomon, has been
assumed; since it is very successful in representing the $Q^2$
dependence of the published high-quality data available to date.
This approach is based on Vector Meson Dominance (VMD) together with
coupling to the continuum which yields widths for the vector mesons
and with asymptotics devised in such a way that the high-$Q^2$
behavior of pQCD is attained for very high momentum transfers ---
just how high is determined by the fit made to the data. No attempt
has been made to refit the model to the most recent experimental
results.  Rather the model is taken to be ``frozen" in the form in
which it was presented in 2005 and thus the excellent agreement with
more recent data may be taken as a test of its predictive power. The
model is summarized in some detail in Sect.~\ref{sec:VMD} together
with discussions of which specific data were fit and the fit results
were presented in Sect.~\ref{sec:momspace}.

In Sect.~\ref{subsec:insights} this reference model has been used to
gain some insights into how the various contributions contained in
it yield the observed behavior of the form factors. Specifically, it
is shown in some detail how having a dipole form for a form factor
is not natural in this approach, but rather arises from compensating
effects where the more natural monopole form factors conspire
effectively to yield roughly the dipole behaviors of the magnetic
form factors at least at modest values of $Q^2$. Such compensations
do not occur for the electric form factor of the proton, in accord
with the data where the ratio $G_E^p/[G_M^p / \mu_p]$ falls with
$Q^2$. All of the ingredients in the GKex model are displayed in
some detail to ascertain which mesons are dominant and which are
less important, at least for modest momentum transfers. Also, the
effects arising from the inclusion of coupling to the continuum (in
this model, only in the $\rho$ meson contributions) are explored by
comparing the form factors obtained with the width present or with
only the $\rho$ pole: these do not differ very significantly,
indicating the relatively minor role played by such effects.

Using the GKex model as a basis the differences between it and the
data have been analyzed using sets of orthonormal functions to
assess the level of uncertainty in the experimental results. In
Sect.~\ref{sec:coord} both the data for the electric form factors
with their uncertainties and the model for these quantities are
Fourier transformed to coordinate space, obtaining the so-called
Breit-frame distribution. It has been emphasized in the discussions
in the body of the paper that, although these are well-defined
mathematically, such Fourier transforms should not be interpreted as
charge distributions. One might ask what use they are, given this
statement. The point of view taken in the present study is that when
one Fourier transforms both the data and the model form factors new
insights into the roles played by the various mesons emerge.
Specifically, it is clearly seen that at large distances ({\it i.e.}
for large Breit-frame Fourier components) the $\rho$ and the
$\omega$ are dominant. As in momentum space, the width of the former
may be turned on or off; the result is only a minor change,
indicating that coupling to the continuum is not a major effect, at
least for such Fourier components. In addition to obtaining the
Breit-frame distributions as discussed above, in the same section
the isoscalar/isovector and u-quark/d-quark distributions are also
extracted for completeness.

The worldwide program over the last two decades to determine the
elastic nucleon form factors using high duty factor electron
accelerators to measure precisely polarization observables has
been highly successful.  It has yielded a data set of
unprecedented precision and consistency for the nucleon elastic
form factors at low and medium $Q^2$.  Although the BLAST
low-$Q^2$ polarized data constitute a very small part of the whole
data set, they have cast doubt on indications seen in earlier data
of structure at this low momentum transfer.  These were attributed
to a ``pion cloud". Such structure is not present in the GKex
representation, and indeed the coupling to explicit continuum
pions is a relatively minor effect in this model, as discussed in
the body of the paper. Further, very high quality measurements at
low-$Q^2$ may help in reaching a definitive answer to the question
of how much structure is actually present.

In this paper, we have used the vector meson dominance model and
this new data set to understand the role of mesons in the
electromagnetic form factors of the proton and neutron. Studies in
both momentum space (for all four form factors) and in coordinate
space (for the Breit-frame distributions that come from the
nucleon's electric form factors) have yielded valuable insights.
In a forthcoming paper, the study will be extended to include new
data for the nucleon magnetic form factors and to investigate the
corresponding coordinate space Breit-frame distributions.


\begin{acknowledgments}

This work has been supported by the United States Department of Energy
under Cooperative Agreement DE-FC02-94ER40818 and the National Science
Foundation Award No. PHY-0855584.
\end{acknowledgments}



\end{document}